\PassOptionsToPackage{unicode}{hyperref}
\PassOptionsToPackage{hyphens}{url}
\PassOptionsToPackage{dvipsnames,svgnames,x11names}{xcolor}
\documentclass[12pt]{article}

\usepackage{amsmath,amssymb}
\usepackage{iftex}
\ifPDFTeX
  \usepackage[T1]{fontenc}
  \usepackage[utf8]{inputenc}
  \usepackage{textcomp} % provide euro and other symbols
\else % if luatex or xetex
  \usepackage{unicode-math}
  \defaultfontfeatures{Scale=MatchLowercase}
  \defaultfontfeatures[\rmfamily]{Ligatures=TeX,Scale=1}
\fi
\usepackage{lmodern}
\ifPDFTeX\else  
\fi
\IfFileExists{upquote.sty}{\usepackage{upquote}}{}
\IfFileExists{microtype.sty}{% use microtype if available
  \usepackage[]{microtype}
  \UseMicrotypeSet[protrusion]{basicmath} % disable protrusion for tt fonts
}{}
\makeatletter
\@ifundefined{KOMAClassName}{% if non-KOMA class
  \IfFileExists{parskip.sty}{%
    \usepackage{parskip}
  }{% else
    \setlength{\parindent}{0pt}
    \setlength{\parskip}{6pt plus 2pt minus 1pt}}
}{% if KOMA class
  \KOMAoptions{parskip=half}}
\makeatother
\usepackage{xcolor}
\makeatletter
\ifx\paragraph\undefined\else
  \let\oldparagraph\paragraph
  \renewcommand{\paragraph}{
    \@ifstar
      \xxxParagraphStar
      \xxxParagraphNoStar
  }
  \newcommand{\xxxParagraphStar}[1]{\oldparagraph*{#1}\mbox{}}
  \newcommand{\xxxParagraphNoStar}[1]{\oldparagraph{#1}\mbox{}}
\fi
\ifx\subparagraph\undefined\else
  \let\oldsubparagraph\subparagraph
  \renewcommand{\subparagraph}{
    \@ifstar
      \xxxSubParagraphStar
      \xxxSubParagraphNoStar
  }
  \newcommand{\xxxSubParagraphStar}[1]{\oldsubparagraph*{#1}\mbox{}}
  \newcommand{\xxxSubParagraphNoStar}[1]{\oldsubparagraph{#1}\mbox{}}
\fi
\makeatother

\usepackage{longtable,booktabs,array}
\usepackage{calc} % for calculating minipage widths
\usepackage{etoolbox}
\makeatletter
\patchcmd\longtable{\par}{\if@noskipsec\mbox{}\fi\par}{}{}
\makeatother
\IfFileExists{footnotehyper.sty}{\usepackage{footnotehyper}}{\usepackage{footnote}}
\makesavenoteenv{longtable}
\usepackage{graphicx}
\makeatletter
\def\maxwidth{\ifdim\Gin@nat@width>\linewidth\linewidth\else\Gin@nat@width\fi}
\def\maxheight{\ifdim\Gin@nat@height>\textheight\textheight\else\Gin@nat@height\fi}
\makeatother
\setkeys{Gin}{width=\maxwidth,height=\maxheight,keepaspectratio}
\makeatletter
\def\fps@figure{htbp}
\makeatother

\makeatletter
\@ifpackageloaded{caption}{}{\usepackage{caption}}
\AtBeginDocument{%
\ifdefined\contentsname
  \renewcommand*\contentsname{Table of contents}
\else
  \newcommand\contentsname{Table of contents}
\fi
\ifdefined\listfigurename
  \renewcommand*\listfigurename{List of Figures}
\else
  \newcommand\listfigurename{List of Figures}
\fi
\ifdefined\listtablename
  \renewcommand*\listtablename{List of Tables}
\else
  \newcommand\listtablename{List of Tables}
\fi
\ifdefined\figurename
  \renewcommand*\figurename{Figure}
\else
  \newcommand\figurename{Figure}
\fi
\ifdefined\tablename
  \renewcommand*\tablename{Table}
\else
  \newcommand\tablename{Table}
\fi
}
\@ifpackageloaded{float}{}{\usepackage{float}}
\floatstyle{ruled}
\@ifundefined{c@chapter}{\newfloat{codelisting}{h}{lop}}{\newfloat{codelisting}{h}{lop}[chapter]}
\floatname{codelisting}{Listing}

\makeatother
\makeatletter
\@ifpackageloaded{caption}{}{\usepackage{caption}}
\@ifpackageloaded{subcaption}{}{\usepackage{subcaption}}
\makeatother

\ifLuaTeX
  \usepackage{selnolig}  % disable illegal ligatures
\fi
\usepackage[]{natbib}
\usepackage{bookmark}

\IfFileExists{xurl.sty}{\usepackage{xurl}}{} % add URL line breaks if available
\hypersetup{
  pdftitle={Title},
  pdfauthor={Author 1; Author 2},
  pdfkeywords={3 to 6 keywords, that do not appear in the title},
  colorlinks=true,
  linkcolor={blue},
  filecolor={Maroon},
  citecolor={Blue},
  urlcolor={Blue},
  pdfcreator={LaTeX via pandoc}}

\newcommand{\anon}{1}

\usepackage{amsmath,amsthm,amssymb}
\usepackage{graphicx}
\usepackage{authblk}

\graphicspath{{Fig/}}

\newtheorem{theorem}{Theorem}

\usepackage[percent]{overpic}
\usepackage{soul}
\usepackage{placeins}
\usepackage{mdframed}
\usepackage{mathtools}
\usepackage[dvipsnames]{xcolor}
\usepackage{tcolorbox}
\usepackage{caption}
\usepackage{pgffor}
\usepackage{algorithm}
\usepackage{algpseudocode}
\usepackage{doi}
\usepackage{setspace}
\usepackage{makecell}
\usepackage{ifthen}
\usepackage{xstring}
\usepackage{array}
\usepackage{keycommand}
\usepackage{booktabs}
\usepackage{orcidlink}

\usepackage{hyperref}
\hypersetup{
  colorlinks=true,
  allcolors=blue
}

\usepackage{xr-hyper}
\newcommand{\suppref}[1]{%
  \ref{supplementary:#1} of the supplementary material%
}

\let\originalleft\left
\let\originalright\right
\renewcommand{\left}{\mathopen{}\mathclose\bgroup\originalleft}
\renewcommand{\right}{\aftergroup\egroup\originalright}
\newcommand{\set}[1]{{\left\lbrace{#1}\right\rbrace}}
\newcommand{\Set}[2]{{\left\lbrace {#1} ~\middle|~ {#2} \right\rbrace}}

\newcommand{\bigsum}[2]{\sum\limits_{#1}^{#2}}

\newcommand{\ind}[1]{\left[{#1}\right]}

\newcommand{\env}[2]{{\begin{#1} #2 \end{#1}}}
\newcommand{\Max}[1]{\max\limits_{#1}}

\newcommand{\paren}[1]{\left({#1}\right)}
\newcommand{\brac}[1]{\left[{#1}\right]}

\newcommand{\E}{\mathbb{E}}

\newcommand{\Prob}[1]{\Pr\paren{#1}}

\newcommand{\tand}{{\text{ and }}}

\newcommand{\tnot}{{\text{not }}}
\newcommand{\x}{{\bf x}}
\newcommand{\reach}[1]{\text{reach } {#1}}
\newcommand{\stopat}[1]{\text{stop at } {#1}}
\newcommand{\through}{\text{ through }}
\newcommand{\reachable}[2]{
    \StrLen{#1}[\LenI]
    \StrLen{#2}[\LenJ]
    \ifthenelse{\LenI=1 \and \LenJ=1}{
        W_{#1#2}
    }{
        W_{#1,#2}
    }
}

\newcommand{\problemframeaftertitle}[1]{
    \ifthenelse{\equal{#1}{}} {} {
        \hfill(\ref*{#1})
    }
}

\newcommand{\mayberefstepcounter}[1]{
    \ifthenelse{\equal{#1}{}} {} {
        \refstepcounter{#1}
    }
}

\newcommand{\maybelabel}[1]{
    \ifthenelse{\equal{#1}{}} {} {
        \label{#1}
    }
}

\newkeycommand+[\|]{\problemframe}[title,counter,label][1]{
    |\begin{tcolorbox}|[
        title=\commandkey{title},
        colback=blue!5!white,
        colframe=blue!75!black,
        after title=|\problemframeaftertitle{\commandkey{label}}|
    ]
    |\mayberefstepcounter{\commandkey{counter}}|
    |\maybelabel{\commandkey{label}}|
    #1
    |\end{tcolorbox}|
}

\newcounter{i}\newcounter{j}

\makeatletter
\newtoks\@tabtoks
\newcommand\addtabtoks[1]{\@tabtoks\expandafter{\the\@tabtoks#1}}
\newcommand*\resettabtoks{\@tabtoks{}}
\newcommand*\printtabtoks{\the\@tabtoks}
\makeatother

\DeclareMathOperator*{\minimize}{minimize\!}

\newtheorem{lemma}[theorem]{Lemma}

\begin{document}

\def\spacingset#1{\renewcommand{\baselinestretch}%
{#1}\small\normalsize} \spacingset{1}

%%%%%%%%%%%%%%%%%%%%%%%%%%%%%%%%%%%%%%%%%%%%%%%%%%%%%%%%%%%%%%%%%%%%%%%%%%%%%%

\if1\anon
{
  \title{\bf Optimal Sequential Testing for Binary Ensemble Classifiers}
  \author{Joseph Kalman \href{mailto:joezkal@gmail.com}{\small \texttt{joezkal@gmail.com}}\thanks{
    We thank Yohai Bar Sinai, Saharon Rosset and David Steinberg for their helpful suggestions.
    AM is supported in part by ISF Grant No. 1662/22 and NSF-BSF Grant No. 2022778.
    Large Language Models (GPT-3.5, -4, o1, -4.5, -4.1, o3, -5, -5.1--5.4; all Claude models up to Opus 4.7; Gemini 2.0 Flash, 2.0 Pro, 2.5 Flash, 2.5 Pro, 2.5 Flash Lite, 3 Pro, 3 Deep Think, 3 Flash, 3.1 Pro)  were used for copyediting this manuscript, code completion, literature search and keyword selection.}\hspace{.2cm}\\
    Department of Statistics and Operations Research, Tel Aviv University \\
    and \\
    Amit Moscovich \href{mailto:mosco@tauex.tau.ac.il}{\small \texttt{mosco@tauex.tau.ac.il}} \\
    Department of Statistics and Operations Research, Tel Aviv University}
  \maketitle
} \fi

\if0\anon
{
  \bigskip
  \bigskip
  \bigskip
  \begin{center}
    {\LARGE\bf Optimal Sequential Testing for Binary Ensemble Classifiers}
\end{center}
  \medskip
} \fi

\bigskip
\begin{abstract}
Ensemble classifiers are predictive models that combine the results of simpler base models, often by majority vote.
A classic example is random forests, which combine the predictions of decision trees.
Ensembles that use more base models can be more accurate but also more costly to train and run.
In this paper, we consider strategies for reducing the computational cost of binary classification using an approach from the field of sequential testing.
Rather than evaluating all the base models and taking a majority vote, we evaluate the base models sequentially and stop execution when a clear majority emerges.
We consider three different notions of optimality for early-stopping strategies that minimize the number of base models executed while controlling the rate of disagreement with the full ensemble.
For each notion of optimality and allowable disagreement rate, we show that a linear program can be constructed and solved efficiently to find the optimal stopping strategy.
We tested these methods on real-world datasets taken from the UC Irvine Machine Learning repository, and on the benchmark datasets proposed by Grinsztajn et al. We found that on most datasets, these methods provide speed-ups of 4x or more while controlling disagreement at 0.1\%.
\end{abstract}

\noindent%
\vfill

\newpage
\spacingset{1.8} % DON'T change the spacing!

%% NON-TEMPLATE CONTENT STARTS HERE

\begin{figure}
    \centering
    \begin{minipage}{\dimexpr\textwidth - 1.9cm\relax}    
        \includegraphics[width=\textwidth, trim=0.8cm 0.8cm 3cm 0cm, clip]{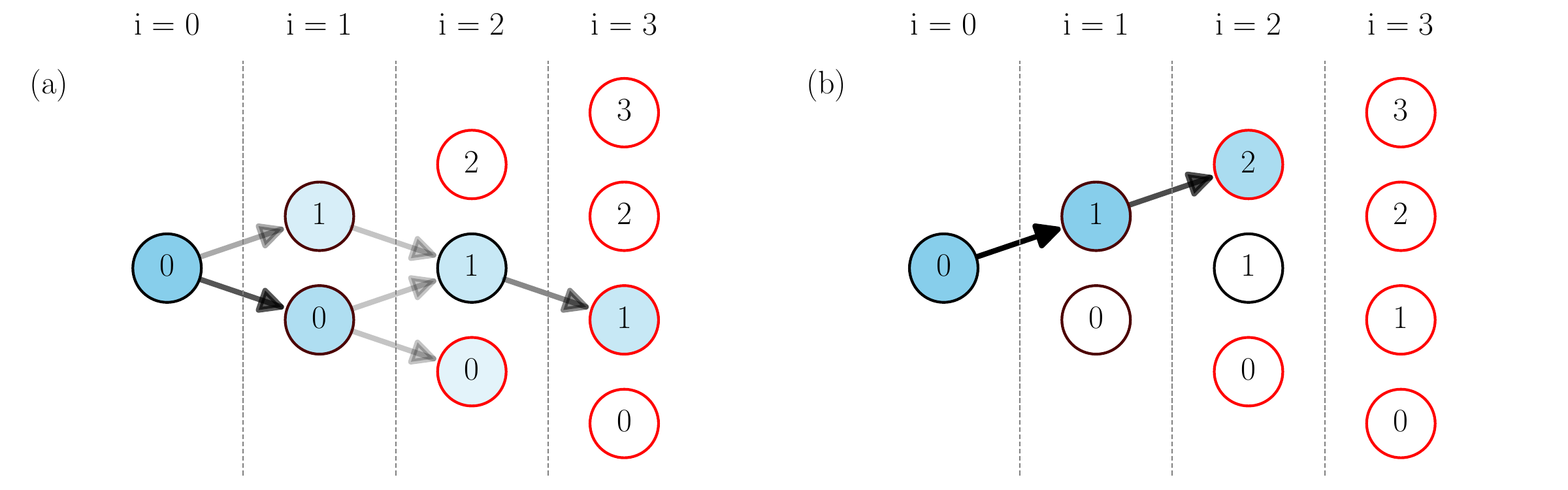}    
        \caption{(a) State transition graph of a sequential test using a certain stopping strategy for $N = 3$ when $n = 1$. The horizontal axis represents $i$, the number of base models executed thus far; the vertical axis and node labels represent $j$, the number of those which have returned positive. The shading of each node or arrow represents the (unconditional) probability of the state or transition occurring. Red node borders represent states where the ensemble will stop if reached.
        (b) The equivalent graph when $n = 3$. Note that only one chain of states is now possible, and testing will never continue past $i = 2$ because by that point there will certainly be 2 positive votes, constituting a majority of the entire ensemble.}
        \label{fig:fwss}
    \end{minipage}
\end{figure}

\section{Introduction}

A binary classifier is a function $f: \mathcal X \to \{ -1, +1 \}$ that assigns elements from a set $\mathcal{X}$ to one of two categories.
An \emph{ensemble classifier} combines the results of multiple base classifiers. One popular example is random forests~\citep{RandomForests}, which combine multiple decision-tree classifiers by majority voting:
\begin{align} \label{eq:majority_classifier}
    f(\x) := \text{Majority}\{f_1(\x), \ldots, f_N(\x)\}.
\end{align}
Ensemble models are useful in an enormous variety of applications, and tree-based ensembles obtain top marks for many kinds of tabular data \citep{RandomForestsBetterThanDL, shwartz-ziv_tabular_2022}.
Choosing the size of the model presents a tradeoff: larger ensembles are more accurate but take longer to train and run. Constraints that apply at testing time can be a concern when the model is to be run on low-power devices or very frequently (e.g., spam filters) \citep{CostSensitiveTreeOfClassifiers}. In this paper, we present a method to sharply reduce the test time of large ensembles using a sequential testing approach which we illustrate in the following example:

Suppose we have trained a random forest with 101 trees to solve a binary classification problem. At test time, we execute one tree at a time and---once all have been executed---we take the majority result.
If it so happens that the first 51 trees all return a positive classification, then the result of the majority vote must be positive and the execution can be stopped at that point.
Even if, having executed only 20 trees, 19 of them have returned a positive result, we can still posit that the ensemble is likely to return a positive label. Thus, by stopping early we can save time at the cost of occasionally disagreeing with the full ensemble.
The example above prompts our central question:
\begin{align}
    \begin{minipage}{0.9\linewidth}
        \begin{quote}
            \emph{What is the speed-up one can achieve for a given disagreement rate using early stopping, and how?}
        \end{quote}
    \end{minipage}
    \label{central-question}
\end{align}
In this paper, ``speed-up" is measured in base models executed, and ``disagreement rate" is the probability of disagreement with the complete ensemble. The answer to this question is a \emph{stopping strategy} that tells us, for each possible state during the execution of the ensemble, whether we should stop and return the current majority. A state, in this context, is a pair $(i, j)$, where $i$ is the number of base models executed so far, and $j$ is the number of base models that returned positive.

Figure~\ref{fig:fwss} shows state transition graphs for an ensemble with $N = 3$ base models. Panel (a) shows possible state transitions when classifying an input for which $n = 1$ base models return positive, whereas in panel (b) all $n=3$ base models return positive. In both cases, execution starts in the state $(0, 0)$. From there, its progress may be nondeterministic due to our assumption that the base models are executed in a random order. 
As this toy example demonstrates, the same stopping strategy may produce a different distribution of state paths depending on the number of positive base models in the ensemble, $n$. Since this number is typically unknown before the ensemble is executed, we must somehow account for all possible values of~$n$. In this paper we present three approaches to this conundrum: a \emph{minimax} worst-case approach, a \emph{minimean} expected-case approach, and a \emph{minimixed} approach which combines the two.

\subsection{Our Contribution}

In this paper we develop several methods for constructing early stopping strategies for ensemble binary classifiers of the form \eqref{eq:majority_classifier} such as random forests.
To do this, we pose the search for a stopping strategy as a minimization problem of the expected stopping time, subject to a constraint that the majority vote of the stopped ensemble must agree with that of the original ensemble with high probability.
Our approach is applicable to any binary classifier based on a majority vote.
The only assumption we make is that the base models are executed in a random order, which simplifies the early-stopping problem, but does exclude ensembles where base models must be executed in a certain order. On the other hand, we avoid other assumptions that have been made in prior work, such as independence of base models.

In Section~\ref{section:Theory}, we present the theory underlying our methods, formalizing the notion of an early-stopping strategy and constructing linear optimization problems whose solution yields optimal stopping strategies. In Section~\ref{section:Computation} we synthesize this theory into a concrete algorithm and discuss computation time and implementation details. In Section~\ref{section:EmpiricalResults}, we present performance metrics for random forests using our methods on a variety of datasets. We end with a discussion of applications and directions for further research in Section~\ref{section:Discussion}.

\subsection{Related Work} \label{section:RelatedWork}

Early stopping is closely related to sequential testing, a field of classical statistics wherein a null hypothesis is tested against an alternative hypothesis by analyzing each observation in sequence and stopping as soon as a conclusion can be reliably drawn \citep{wald_sequential_1945, Siegmund1985}.
It is commonly used in contexts such as quality control and medical trials where reaching conclusions as early as possible is key.
Our problem is closely related to the special case of sequential testing with binary outcomes, for which \citet{SilvaKulldorffKatherineYih2020} present a solution that is provably optimal in the sense that interests us. However, their work is not directly applicable to our problem because their method assumes independence of the base models, an assumption that we do not wish to make; for further discussion see Section~\suppref{section:BinomialVsHypergeometric}. \citet{Lai1979} takes a very similar approach to us---sequential testing on binary data for finite populations, using a hypergeometric model---but does not present a strategy that is provably optimal.

Our central example of an ensemble model, the random forest, was first presented in its modern form by \citet{RandomForests}. Since then, several approaches to reduce their prediction time were suggested. One popular approach is to prune the forest by either deleting individual branches or entire trees during training \citep{ManzaliElfar2023, surjanovic_alpha-trimming_2024}.
There are several works that consider early-stopping strategies for ensemble models: the method of \citet{QuitWhenYouCan} is based on dynamically reordering the base models during testing, combined with an early-stopping strategy.
\citet{fan_pruning_2002} present pruning, dynamic-ordering and early-stopping techniques specifically for scenarios with asymmetric loss.
An approach to early-stopping that is similar to ours was presented by  \citet{SchwingZachZhengPollefeys2011}, but it is based on binomial confidence bands under the simplifying assumption that base models are independent of each other, an assumption that does not hold for random forests. Another recent work is that of \citet{DagheroBurrelloMaciiMontuschiEtAl2023}, where they consider a heuristic approach to early stopping with no theoretical guarantees.
To the best of our knowledge, our method is the first to achieve provably-optimal results on majority-based binary ensemble classifiers, without making further assumptions such as independence.
The only assumption that we make is that the base models $f_1({\x}),\, f_2({\x}),\,  \dots\,, f_N({\x})$ are executed in a random order.

\section{Theory} \label{section:Theory}

In this section, we formalize the central question \eqref{central-question} posed in the introduction, as follows:
first, in Section~\ref{section:StoppingStrategiesAndTheirEvaluation}, we formally define the notion of a stopping strategy and motivate the metrics we use to compare strategies. Next, in Section~\ref{section:ComputingProbStopAt}, we show how these metrics can be computed.
We then describe several optimization problems in Section~\ref{section:OptimizationProblems}, each of which finds the stopping strategy which is optimal in a different sense. Finally, in Section~\ref{section:LinearProgramming}, we reformulate them as linear programs that can be efficiently solved.
For the reader's convenience, common symbols and terms used in this Section are summarized in Table~\ref{tab:symbols}.

\begin{table}[!t]
    \vspace{0.5ex}
    \caption{Glossary of symbols and terms.  \label{tab:symbols}}
    \vspace{-0.5ex}
    \renewcommand{\arraystretch}{1.3}
    \begin{tabular*}{\textwidth}{>{\centering\arraybackslash}p{0.25\linewidth}p{0.70\linewidth}}
        \toprule
        {\bf Symbol} & {\bf Meaning} \\
        \midrule
        $N$ & Total number of base models in the ensemble \\
        $n$ & Number of base models that would return a positive classification on the current input \\
        $D$ & True probability distribution of $n$, given a trained ensemble \\
        $\hat{D}$ & Estimated or assumed probability distribution of $n$ \\
        state $(i, j)$ & State wherein $i$ base models have been executed and $j$ have returned a positive classification \\
        $\theta$ & Stopping strategy: on reaching a state $(i, j)$, the ensemble will stop with probability $\theta_{ij}$ \\
        $\Theta_N$ & The set of valid stopping strategies for ensembles of size $N$ \\
        $B$ & Number of base models executed before stopping \\
        disagreement rate, $\Prob{\text{disagree}}$ & Probability that the early-stopped ensemble gives a different result than the full ensemble \\
        ADR, $\alpha$ & Allowable Disagreement Rate, the user-defined maximum disagreement rate  \\
        $\reach{(i, j)}$ & Event that the ensemble enters the state $(i, j)$ while running, whether or not it stops there \\
        $\stopat{(i, j)}$ & Event that the ensemble enters the state $(i, j)$ while running and stops there \\
        $\reachable i j$ & Event that the ensemble \textit{would} enter the state $(i, j)$ if not stopped; equivalently, that precisely $j$ of the first $i$ base models are positive \\
        $d_{ij}(N, n)$ & Indicator of whether the state $(i, j)$ is a disagreement state \\
        $p, \pi, \bar \pi$ & Decision variables representing conditional probabilities induced by $\theta$ (see \eqref{pi-interp}) \\
        $\Pi$ & Set of valid values for $(p, \pi, \bar \pi)$ (see \eqref{formula:big-pi}) \\
        $\ind{\textit{expression}}$ & Iverson bracket; evaluates to $1$ if \textit{expression} is true, otherwise to $0$ \\
        \bottomrule
    \end{tabular*}
\end{table}

\subsection{Stopping Strategies and Their Evaluation} \label{section:StoppingStrategiesAndTheirEvaluation}

Consider a binary ensemble classifier with $N$ base models that are executed in a random order. A \emph{deterministic stopping strategy} is a boolean matrix $\theta \in \set{0, 1}^{(N + 1) \times (N + 1)}$, indexed from zero in both dimensions. The entry $\theta_{ij}$ tells us what to do if, during testing, we reach a state where precisely $i \in \{0, \ldots, N \}$ base models have been executed and precisely $j$ of them have returned positive: if $\theta_{ij} = 1$ we stop early and return the majority result so far $\ind{j > i / 2}$, and if $\theta_{ij} = 0$, we continue executing base models.
A more general (nondeterministic) \emph{stopping strategy} is a matrix of real values $\theta \in [0, 1]^{(N + 1) \times (N + 1)}$, still indexed from zero. We interpret it the same way, except that now $\theta_{ij}$ tells us the \textit{probability} with which we should stop if we reach the relevant state. Since testing \textit{must} stop once the ensemble is exhausted, the set of valid stopping strategies $\Theta_N$ includes only those for which $\theta_{Nj} = 1$ for all $j$.

Given a stopping strategy and a certain input $\x$ being classified, where will we actually stop? The answer to this is a probability distribution, both because the stopping strategy itself may be nondeterministic and because the order in which the base models are executed is random. We will soon show how we can compute the probability of stopping at any given state $(i, j)$, conditional on the ensemble size $N$, the number $n$ of base models returning positive, and the stopping strategy $\theta \in \Theta_N$. We denote this as $\Prob{\stopat{(i, j)} \mid N, n, \theta}$. Note that we need not condition on $\bf x$ because $\bf x$ affects the stopping behavior only through $n$.
From these probabilities, we can compute two metrics that are of interest in the evaluation of stopping strategies. First is the number of base models executed before stopping, denoted $B$, or rather its expectation $\E[B \mid N, n, \theta]$. Since the number of base models executed so far is the same as the step number $i$, this expectation is given by
\env{align}{
    \E\brac{B \mid N, n, \theta}
    =&\; \bigsum{i=0}{N} i \cdot \Prob{\stopat{\text{step } i} \mid N, n, \theta} \nonumber \\ 
    =&\; \bigsum{i=0}{N} i \cdot \bigsum{j=0}{i} \Prob{\stopat{(i, j)} \mid N, n, \theta}.
    \label{formula:expected-B}
}
The second metric of interest is the \emph{disagreement rate}: the probability of the sequentially-tested ensemble yielding a different result from that of the complete ensemble. This occurs if and only if we stop in a ``disagreement state", that is, a state where the majority result so far is different from that of the ensemble as a whole. We indicate such states using the Boolean matrix $d(N, n)$, defined as
\env{align}{
    d_{ij}(N, n) := \ind{j > i / 2} \neq \ind{n > N / 2},
}
\noindent where $\ind{\cdot}$ is the Iverson bracket; for the purpose of our analysis we assume that tied votes yield a negative prediction. Using this notation, the disagreement rate is
\env{align}{
    \Prob{\text{disagree} \mid N, n, \theta}
    = \bigsum{i=0}{N} \bigsum{j=0}{i} \Prob{\stopat{(i, j)} \mid N, n, \theta} d_{ij}(N, n).
    \label{formula:prob-disagree}
}

Our decision to minimize disagreement rate rather than error rate comes from the observation that the error rate of the early-stopped ensemble is highly dependent on that of the complete ensemble, over which we have no control. Intuitively, however, we would not expect early stopping to reduce the error rate in the general case, but rather to increase it. The disagreement rate acts as a bound on the excess error rate introduced by early stopping, i.e.,
\env{align}{
    \Prob{\text{error with early-stopping}} \leq \Prob{\text{error with full ensemble}} + \Prob{\text{disagree}}.
}

\subsection{Computing Conditional Stopping Probabilities} \label{section:ComputingProbStopAt}

We now explain how to compute the probability of stopping at a state $(i, j)$ conditional on $N, n, \theta$. Applying the results in Section~\ref{section:StoppingStrategiesAndTheirEvaluation} will then allow us to express the metrics of interest as a function of these parameters, and thus to formulate our optimization problems in Section~\ref{section:OptimizationProblems}.

For brevity, we leave the conditioning on $N, n, \theta$ implicit in all the probabilities that follow in this subsection. The probability of stopping at a given state is simply
\env{align}{
    \Prob{\stopat{(i, j)}} =&\; \Prob{\reach{(i, j)}} \cdot \theta_{ij}. \label{formula:prob-reach-and-stop}
}
The probability of \textit{reaching} a given state can be computed step-by-step using a recursive formula. The base case is \(\Prob{\reach{(0, 0)}} = 1\).
From each state $(i, j)$, we transition to the state $(i+1, j+1)$ if the next base model returns positive, and to $(i+1, j)$ if it returns negative. Since base models are randomly ordered, the probability of it being one of those that returns positive is
\env{align}{
    \Pr&\paren{(i+1)\text{-th model is positive} \mid \reach {(i, j)}} = \tfrac{n - j}{N - i}.
}
Flipping our perspective, each state $(i, j)$ where $i > 0$ can be reached only through state $(i - 1, j - 1)$ or through the state $(i - 1, j)$. Both of these paths can only occur if we do not stop in the $(i-1)$-th state. Therefore the probability of the first path, wherein \( f_i(\x) \) is positive, is
\env{align}{
        \Pr&\paren{\reach{(i, j)} \through {(i-1, j-1)}} \nonumber \\
        &= \Prob{\reach{(i-1, j-1)}} \cdot (1-\theta_{i-1,j-1}) \cdot \tfrac{n - (j - 1)}{N - (i - 1)},
}
and likewise the probability of the second path, wherein \(f_i(\x)\) is negative, is
\env{align}{
        \Pr&\paren{\reach{(i, j)} \through {(i-1, j)}}  \nonumber \\
        &= \Prob{\reach{(i-1, j)}} \cdot (1-\theta_{i-1,j}) \cdot \paren{1 - \tfrac{n - j}{N - (i - 1)}}.
}
\noindent The total probability of reaching state $(i, j)$ is simply the sum of these two terms,
\env{align}{
    \Pr&\paren{\reach{(i, j)}} \nonumber \\
    &= \Pr\paren{\reach{(i, j)} \through {(i-1, j)}} \nonumber \\
    &\;+ \Pr\paren{\reach{(i, j)} \through {(i-1, j-1)}} \nonumber \\
    &= \Prob{\reach{(i-1, j-1)}} \cdot (1-\theta_{i-1,j-1}) \cdot \tfrac{n - (j - 1)}{N - (i - 1)} \nonumber \\
    &\;+ \Prob{\reach{(i-1, j)}} \cdot (1-\theta_{i-1,j}) \cdot \paren{1 - \tfrac{n - j}{N - (i - 1)}}. \label{formula:prob-reach-recursive-step-full}
}
Combining $\eqref{formula:prob-reach-recursive-step-full}$ with the base case above ($\Prob{\reach{(0, 0)}} = 1$) allows us to compute $\Prob{\reach{(i, j)}}$ for any $(i, j)$, from which  $\Prob{\stopat{(i, j)}}$ is derived for all $i,j$ via Eq.~\eqref{formula:prob-reach-and-stop}.
Finally, by plugging $\Prob{\stopat{(i, j)}}$ into equations \eqref{formula:expected-B} and \eqref{formula:prob-disagree}, we obtain the expected runtime and disagreement rate of the stopping strategy $\theta$, given $n$.

\subsection{Finding Optimal Stopping Strategies} \label{section:OptimizationProblems}

Having expressed the disagreement rate and expected runtime as a function of $N, n, \theta$, we can now express our optimization problem: we wish to find the stopping strategy which minimizes the expected runtime while keeping the disagreement rate below a specified threshold $\alpha$, i.e.,
\env{align}{
    \minimize_{\theta \in \Theta_N}
    \ \E[B]
    \quad
    \text{s.t.}
    \quad
    \Pr(\text{disagree}) \leq \alpha
}
\noindent or, more explicitly,
\env{align}{
    \minimize_{\theta \in \Theta_N}
    \ \E[B \mid N, n, \theta]
    \quad
    \text{s.t.}
    \quad
    \Pr(\text{disagree} \mid N, n, \theta) \leq \alpha.
}
But this problem is not immediately solvable, because we don't know $n$; therefore, we must aggregate over \emph{possible} values of $n$. We can do this either by considering worst-case values of $n$ or---if we are willing to assume a distribution on $n$---by taking an expectation. We consider three approaches to finding optimal stopping strategies. First, we have the \emph{minimax approach}, where we minimize the worst-case expected runtime among all strategies whose worst-case disagreement rate is at most $\alpha$:

\problemframe[title={Problem: $\Theta$-Minimax}, counter=problem, label={problem:theta-minimax}]{
    \vspace{-30pt}
    \env{align}{
        \minimize_{\theta \in \Theta_N}
        \ \max_n
        \ \E[B \mid N, n, \theta]
        \qquad
        \text{s.t.}
        \qquad
        \max_n
        \Pr(\text{disagree} \mid N, n, \theta) \leq \alpha. \nonumber
    }
}
Next, we have the \emph{minimean approach}, where we estimate or assume some distribution $\hat D$ for $n$ and minimize the overall expected runtime among all strategies whose expected disagreement rate is below $\alpha$:

\problemframe[title={Problem: $\Theta$-Minimean}, counter=problem, label={problem:theta-minimean}]{
    \vspace{-30pt}
    \env{align}{
        \minimize_{\theta \in \Theta_N}
        \ \E_{n \sim \hat{D}} \brac{
            \E[B \mid N, n, \theta]
        }
        \qquad
        \text{s.t.}
        \qquad
        \E_{n \sim \hat{D}} \brac{
            \Pr(\text{disagree} \mid N, n, \theta)
        }  \leq \alpha. \nonumber
    }
}
Finally, we have the \emph{minimixed approach}, where we assume some distribution for~$n$ and minimize the overall expected runtime among all strategies whose worst-case disagreement rate is below $\alpha$:
\problemframe[title={Problem: $\Theta$-Minimixed}, counter=problem, label={problem:theta-minimixed}]{
    \vspace{-30pt}
    \env{align}{
        \minimize_{\theta \in \Theta_N}
        \ \E_{n \sim \hat{D}} \brac{
            \E[B \mid N, n, \theta]
        }
        \quad
        \text{s.t.}
        \quad
        \max_n
        \Pr(\text{disagree} \mid N, n, \theta) \leq \alpha. \nonumber
    }
}
As discussed in Section~\ref{section:StoppingStrategiesAndTheirEvaluation}, the quantities $\E[B\,{\mid}\,N, n, \theta]$ and $\Pr(\text{disagree} \,{\mid}\, N, n, \theta)$ in our optimization problems are both linear combinations of $\Prob{\stopat{(i, j)}}$ for various values of $(i, j)$. These probabilities are in turn easily computable from the formulae in Section~\ref{section:ComputingProbStopAt}. 
However, our expression for $\Prob{\stopat{(i, j)}}$ in Equation~\eqref{formula:prob-reach-and-stop} depends on that for $\Prob{\reach{(i, j)}}$, which in turn is expressed using a recursive formula in Equation~\eqref{formula:prob-reach-recursive-step-full}, so $\Prob{\stopat{(i, j)}}$ involves the product of multiple decision variables. Thus Problems~\eqref{problem:theta-minimax}, \eqref{problem:theta-minimean}, and \eqref{problem:theta-minimixed} are not linear programs, nor even quadratically-constrained quadratic programs (QCQPs). However, they can be reformulated as QCQPs with $O(N^3)$ decision variables.
Unfortunately, in this formulation the problems are impractical to solve even for $N \approx 16$ (for more on this reformulation and our experiments, see Section~\suppref{section:SolutionTimes}).
In the next subsection we explain how these optimization problems can be solved efficiently by reformulating them as linear programs.

\subsection{Optimization by Linear Programming} \label{section:LinearProgramming}

Real ensembles often have hundreds of base models, so a more computationally-tractable approach is necessary. For this purpose we use a technique from \citet{SilvaKulldorffKatherineYih2020}. Essentially, we reformulate the problems in terms of a new set of decision variables representing conditional probabilities induced by $\theta$. With these decision variables, Problems \eqref{problem:theta-minimax}, \eqref{problem:theta-minimean}, and \eqref{problem:theta-minimixed} can easily be expressed as linear programs and solved with known efficient techniques \citep{HillierLieberman2024}. Once the linear program has been solved, the optimal values of $\theta$ can be easily extracted from those of the new decision variables.
We call the new decision variables $p_{ij}, \pi_{ij}, \bar \pi_{ij}$, and assign them the following meanings:
\env{align}{
    p_{ij} &:= \Prob{\reach{(i, j)} \mid N, n, \theta, \reachable i j} \nonumber \\
    \pi_{ij} &:= \Prob{\stopat{(i, j)} \mid N, n, \theta, \reachable i j}  \label{pi-interp} \\
    \bar \pi_{ij} &:= \Prob{\reach{(i, j)} \tand \tnot \stopat{(i, j)} \mid N, n, \theta, \reachable i j}, \nonumber
}

\noindent where $\reachable i j$ is the event that the state $(i, j)$ \textit{would} be reached if no stopping strategy were used---that is, that precisely $j$ of the first $i$ base models are positive. Conditioning on this event as well as $n$ makes the value of these conditional probabilities the same for all $n$, which is why they are indexed by $(i, j)$ rather than $(i, j, n)$. Note that $W_{ij}$ has probability zero whenever $j > i$; we adopt the convention that conditioning on an event of probability zero yields a probability of zero, so in those cases $p_{ij} = \pi_{ij} = \bar \pi_{ij} = 0$. The meanings of the decision variables imply certain conditions; we define $\Pi$ as the set of triplets satisfying these conditions, i.e.:
\problemframe[title={Definition of $\Pi$}]{
    \vspace{-30pt}
    \env{align}{
        \Pi :=& \Set{(p, \pi, \bar \pi) \in \paren{[0, 1]^{(N+1) \times (N+1)}}^3}{\text{\eqref{sky-condition-first-step}--\eqref{sky-condition-last-step} hold}} \label{formula:big-pi} \\
        p_{00} =&\; 1 \tag{\theequation.1} \label{sky-condition-first-step} \\
        p_{ij} =&\; \pi_{ij} = \bar \pi_{ij} = 0 \text{ for } i = 0, \dots, N, ~ j = i + 1, \dots, N \tag{\theequation.2} \label{sky-condition-unreachable} \\
        \pi_{ij} + \bar \pi_{ij} =&\; p_{ij} \text{ for } i = 0, \dots, N,~ j = 0, \dots, i \tag{\theequation.3} \label{sky-condition-pi-plus-pi-bar} \\
        p_{i + 1, j + 1} =&\; \frac{i - j}{i + 1} \bar \pi_{i, j + 1} + \frac{j + 1}{i + 1} \bar \pi_{ij} \text{ for } i = 0, \dots, N - 1,~ j = 0, \dots, i \tag{\theequation.4} \label{sky-condition-combinatoric} \\
        \pi_{Nj} =&\; p_{Nj} \text{ for } j = 0, \dots, N. \tag{\theequation.5} \label{sky-condition-last-step}
    }
}
Conditions~\eqref{sky-condition-first-step}, \eqref{sky-condition-unreachable}, and \eqref{sky-condition-pi-plus-pi-bar} follow directly from \eqref{pi-interp}.
Condition~\eqref{sky-condition-combinatoric} is due to the way each state follows only from its two antecedent states. Condition~\eqref{sky-condition-last-step} is a reformulation of the constraint $\theta_{Nj} = 1$, reflecting the fact testing must halt once the ensemble is exhausted.

The following lemmas, proofs of which may be found in Section~\suppref{section:Proofs}, express the fact that membership in $\Pi$ is both necessary and sufficient for a given $p, \pi, \bar \pi$ to satisfy the definition in $\eqref{pi-interp}$ for some stopping strategy $\theta$.
\begin{lemma}[Necessity]
    \label{claim:exists-pi-for-theta}
    For each $N, n, \theta \in \Theta_N$, if $p, \pi, \bar \pi$ are defined as in Equation~\eqref{pi-interp}, then $(p, \pi, \bar \pi) \in \Pi$.
\end{lemma}
\begin{lemma}[Sufficiency]
\label{claim:can-find-theta-for-pi}
    Fix some $N, n$. For any $(p, \pi, \bar \pi) \in \Pi$, there exists some stopping strategy $\theta \in \Theta_N$, computable in $O(N^2)$ time from $(p, \pi, \bar \pi)$, such that Equation~\eqref{pi-interp} holds.
\end{lemma}
Next, Equation~\eqref{pi-interp} allows us to express the expected runtime and disagreement rate as linear combinations of $\pi$. Recall that in Section~\ref{section:StoppingStrategiesAndTheirEvaluation} we saw
\env{align}{
    \E[B \mid N, n, \theta] =&\; \bigsum{i=0}{N} i \cdot \bigsum{j=0}{i} \Prob{\stopat{(i, j)}}, \\
    \Prob{\text{disagree} \mid N, n, \theta} =&\; \bigsum{i=0}{N} \bigsum{j=0}{i} \Prob{\stopat{(i, j)}} \ind{\ind{j > i / 2} \neq \ind{n > N / 2}}.
}
The only term in these sums that depends on $\theta$ is $\Prob{\stopat{(i, j)}}$,
\env{align}{
    \Prob{\stopat{(i, j)}}
    =
    \Prob{\stopat{(i, j)} \mid \reachable i j} \Prob{\reachable i j}
    =
    \pi_{ij} \Prob{\reachable i j}. \label{formula:prob-stop-at}
}
\noindent The quantity $\Prob{\reachable i j}$---the probability that precisely $j$ of the first $i$ base models are positive---distributes hypergeometrically, i.e.,
\env{align}{
    \Prob{\reachable i j}
    =\; \tfrac{\binom {n} j \binom {N - n} {i - j}}{\binom N i}\ , \label{formula:prob-reachable}
}
and does not depend on $\theta$. That makes the following a \textit{linear} optimization problem:

\problemframe[title={Problem: $\Pi$-Minimax (reformulation of $\Theta$-Minimax)}, counter=problem, label={problem:pi-minimax}]{
    \vspace{-30pt}
    \env{align}{
        \minimize_{(p, \pi, \bar \pi) \in \Pi}
            \quad \Max{n} \E[B \mid N, n, \theta] \quad
        \text{s.t.} \quad
            \Max{n} \Prob{\text{disagree} \mid N, n, \theta} \leq \alpha \nonumber
    }
}

\noindent where $\theta$ is implicitly given by $(p, \pi, \bar \pi)$: due to the probabilities each variable represents, we have $\theta_{ij} = \pi_{ij} / p_{ij}$ wherever $p_{ij} \neq 0$, and an arbitrary value when $p_{ij} = 0$. Problem~$\Pi$-Minimax~\eqref{problem:pi-minimax} is thus equivalent to Problem~$\Theta$-Minimax~\eqref{problem:theta-minimax} by the following theorem, a proof of which can be found in Section~\suppref{section:Proofs}.
\env{theorem}{\label{claim:minimax-problem-pi-theta-equivalent}
    Problem~$\Theta$-Minimax~\eqref{problem:theta-minimax} is polynomially reducible to Problem~$\Pi$-Minimax~\eqref{problem:pi-minimax} in $O(N^2)$ time.
}
\noindent By an analogous argument, we can reformulate Problem $\Theta$-Minimean~\eqref{problem:theta-minimean} as
\problemframe[title={Problem: $\Pi$-Minimean (reformulation of $\Theta$-Minimean)}, counter=problem, label={problem:pi-minimean}]{
    \vspace{-30pt}
    \env{align}{
        \minimize_{(p, \pi, \bar \pi) \in \Pi}
        \ \E_{n \sim \hat{D}} \brac{
            \E[B \mid N, n, \theta]
        }
        \quad
        \text{s.t.}
        \quad
        \E_{n \sim \hat{D}} \brac{
            \Pr(\text{disagree} \mid N, n, \theta)
        }  \leq \alpha. \nonumber
    }
}
\noindent We can likewise reformulate Problem $\Theta$-Minimixed~\eqref{problem:theta-minimixed} as 
\problemframe[title={Problem: $\Pi$-Minimixed (reformulation of $\Theta$-Minimixed)}, counter=problem, label={problem:pi-minimixed}]{
    \vspace{-30pt}
    \env{align}{
        \minimize_{(p, \pi, \bar \pi) \in \Pi}
        \ \E_{n \sim \hat{D}} \brac{
            \E[B \mid N, n, \theta]
        }
        \quad
        \text{s.t.} \quad
            \Max{n} \Prob{\text{disagree} \mid N, n, \theta} \leq \alpha. \nonumber
    }
}
\noindent We now have a method to compute optimal stopping strategies of three different kinds: by solving Problem~$\Pi$-Minimax~\eqref{problem:pi-minimax}, we can find a stopping strategy that minimizes runtime and guarantees the disagreement rate \textit{in the worst case}; by solving Problem~$\Pi$-Minimean~\eqref{problem:pi-minimean} we can find a stopping strategy that minimizes runtime and guarantees the disagreement rate \textit{in expectation}, provided we have a distribution for $n$; or, by solving Problem~$\Pi$-Minimixed~\eqref{problem:pi-minimixed}, we can find a stopping strategy that minimizes runtime in expectation while guaranteeing the disagreement rate in the worst case.

\section{Computation} \label{section:Computation}

\begin{table}[b]
    \vspace{1ex}
    \caption{{\bf Methods.} Early-stopping approaches analyzed in this paper. Average case is with regard to an assumed or estimated distribution of $n$ (the number of positive votes in the ensemble). We tested the minimean and minimixed approaches both on the empirical distribution of a holdout calibration set (Cal) and on a uniform distribution (Flat).  \label{tab:approaches}}
    %\vspace{1em}
    \renewcommand{\arraystretch}{1.2}
    \begin{tabular*}{\textwidth}{cccc}
        \toprule
        \textbf{Name} & \makecell{\textbf{Optimizes runtime} \\ \textbf{for...}} & \makecell{\textbf{Bounds disagreement} \\ \textbf{rate of...}} & \makecell{\textbf{Formulation}} \\
        \midrule
        Minimax & Worst case & Worst case & Problem~\eqref{problem:pi-minimax} \\
        Minimean & Average case & Average case & Problem~\eqref{problem:pi-minimean} \\
        Minimixed & Average case & Worst case & Problem~\eqref{problem:pi-minimixed} \\
        Schwing et al. & --- & --- & \cite{SchwingZachZhengPollefeys2011} \S 3.1 \\
        \bottomrule
    \end{tabular*}
\end{table}

The theory in Section~\ref{section:Theory} comes together to produce algorithms for finding stopping strategies which are optimal in the minimean, minimax, or minimixed sense (see Table~\ref{tab:approaches}). In this Section, we discuss some computational details. Detailed pseudocode can be found in Section~\suppref{section:Pseudocode}.
A link to our Python implementation can be found in Section~\ref{sec:reproducibility}.

The basic outline of the algorithm is the same regardless of whether we take a minimax, minimean, or minimixed approach: we construct a linear program, solve it, and convert the solution to a stopping strategy. The time complexity of setting up the linear program is $O(N^3)$, and the time complexity of converting the solution to a stopping strategy is $O(N^2)$.
The bulk of the time complexity in finding a stopping strategy is thus in solving the linear program. The most time-efficient algorithm of which we are aware for doing this is that given by \citet{DeterministicLinearProgrammingSolver}, which takes $\tilde O\paren{v^\omega \log (v / \delta)}$ time, where $v$ is the number of decision variables, $\delta$ is the required relative precision, $\omega$ is the exponent of the time complexity of matrix multiplication, and $\tilde O$ hides a poly-log factor.

We used the linear program solver SoPlex~6.0.0, part of the SCIP Optimization Suite 8.0.0 \citep{bestuzheva_2023}. Since the state transition probabilities are all ratios of integers, we used the rational solver options to avoid loss of precision on large ensembles \citep{SoPlexRational1, SoPlexRational2, SoPlexRational3}.
Empirically, solving the optimization problems takes less than a minute up to $N = 100$; see Section~\suppref{section:SolutionTimes} for technical details and solution times.

\section{Empirical Results} \label{section:EmpiricalResults}

\subsection{Computed Stopping Strategies} \label{section:Precomputation}

A key difference between the minimax approach and the other two approaches presented in Section~\ref{section:OptimizationProblems} is that computing a minimax strategy requires us to specify only the size of the ensemble $N$ and the allowable disagreement rate $\alpha$, whereas for a minimean or minimixed strategy we must also supply an estimated or assumed distribution $\hat D$ of the number of positive base models $n$. Therefore, only minimax stopping strategies can be precomputed without regard for the data.

\begin{figure}
    \includegraphics[width=\textwidth]{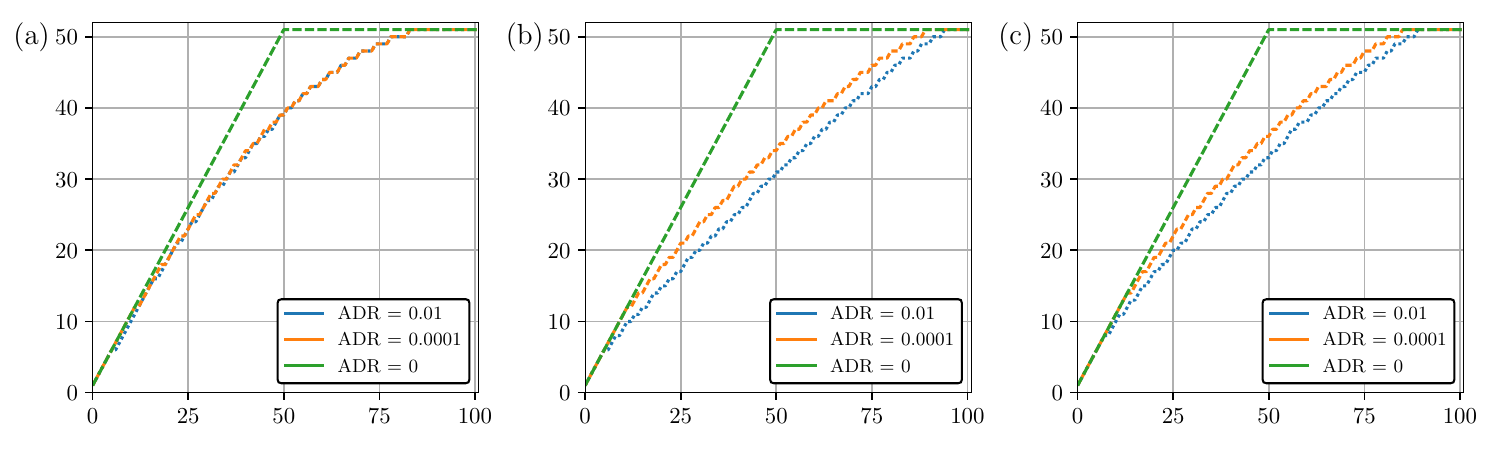}
    \caption{Envelopes of precomputed stopping strategies for ensembles of size $N = 101$ and various values of the ADR (allowable disagreement rate) $\alpha$. The horizontal axis is $i$, the number of base models executed; the vertical axis is $j$, the number of those which were positive. The plotted lines represent $b_i$, the number of positive votes which must be reached to stop.
    {\bf (a)} Minimax {\bf (b)} Minimean {\bf (c)} Minimixed.}
    \label{fig:stopping-strategies}
\end{figure}

We precomputed minimax stopping strategies for all ensemble sizes $11 \leq N \leq 121$, for allowable disagreement rates of $\alpha \in \set{0, 10^{-6}, 10^{-5}, \dots, 10^{-1}}$. We did likewise for minimean and minimixed stopping strategies under the assumption of uniformly distributed $n$. The precomputed strategies are available as part of our GitHub repository.
In practice, all precomputed stopping strategies were almost entirely deterministic---that is, all but two  entries $\theta_{ij}$ were either zero or one. Moreover, they all followed the same simple stopping rule: each strategy had some step $i_0$ before which there was no stopping at all, and after which stopping was certain iff the observed number of positives $j$ was greater than $b_i$ or less than $a_i$ for thresholds $a_i$ and $b_i$ which were different for each stopping strategy. Nondeterminism only emerged at step $i_0$ itself, where the strategy was to stop with some positive probability iff the consensus so far was unanimous. For minimax stopping strategies, and for minimean and minimixed strategies based on a symmetrical distribution of $n$, the strategies themselves were also symmetrical, which is to say they satisfied $b_i - i/2 = i/2 - a_i$ for all $i$.
This simple shape means we lose almost no information by plotting each stopping strategy as a single line giving $b_i$ for each $i$, or a trivial threshold of $i + 1$ for $i < i_0$. We refer to this as the \emph{envelope} of the stopping strategy. Envelopes of selected stopping strategies computed using the minimax, minimean, and minimixed approaches are shown in Figure~\ref{fig:stopping-strategies}. When the ADR is zero, the envelope starts at $i + 1$---in other words, never stop, because $j$ can never exceed $i$ anyway---until we have observed a majority of the entire ensemble ($51$ base models); thereafter, we stop iff we have $51$ votes in the same direction, since then and only then can the result of the whole ensemble be predicted with certainty. The envelopes for higher ADRs are lower, representing more aggressive stopping strategies.

\subsection{Empirical Evaluation}

We tested each early-stopping approach on random forests trained on twenty-seven different real-world datasets, with allowable disagreement rates ranging from $0$ to $0.1$ on each dataset.
The main collection of eight datasets was taken from the UC Irvine Machine Learning Repository \citep{UCIMLRepo}. Datasets were only included if they were tagged with the ``Classification" task, had a column marked with the ``Target" role, had no missing data, were non-synthetic, and were available for import in Python. Among datasets matching these criteria, the eight largest datasets (by number of instances) were selected. The selected datasets, ranging in size from 58k to 581k observations, are detailed in Table~\ref{tab:datasets}. In addition to these, we also ran all experiments on those benchmark datasets proposed by \citet{RandomForestsBetterThanDL} which were marked as appropriate for classification problems, of which there were nineteen; for readability, some figures in the main paper exclude these datasets, but the results were substantially similar to the main eight, and more complete figures appear in Section~\suppref{section:BenchmarkResults}.
\begin{table}[!t]
    \footnotesize
    \vspace{1ex}
    \caption{Datasets used in this paper. ``ID" refers to the dataset's ID in the UCI Machine Learning Repository. {\bf \#} is number of observations, $p$ is number of features, and $C$ is number of classes before dichotomization. ``\%pos" is the percentage belonging to the largest class, which forms the positive class after dichotomization.  \label{tab:datasets}}
    %\vspace{1em}
    \renewcommand{\arraystretch}{1.2}
    \begin{tabular*}{\textwidth}{lllllp{0.54\linewidth}}
        \toprule
        {\bf Name (ID)} & {\bf \#} & {\bf p} & {\bf C} & {\bf \%pos} & {\bf Description} \\
        \midrule
        \makecell[tl]{Ground Cover\\(31)}
        & 581,012 & 54 & 7 & 48.8 & Classification of pixels into forest cover types based on attributes such as elevation, slope, hillshade, soil-type, etc. \\
        Income (117)
        & 199,523 & 41 & 2 & 93.8 & Weighted census data extracted from the 1994 and 1995 current population surveys conducted by the U.S. Census Bureau. Target variable is whether income is $\geq$ \$50k / year. \\
        Diabetes (891)
        & 253,680 & 21 & 2 & 86.1 & Features are demographics, lab test results, and answers to survey questions. Target variable is whether a patient has diabetes, is pre-diabetic, or is healthy.
        \\
        Skin (229)
        & 245,057 & 3 & 2 & 79.2 & RGB color triplets must be classified as skin or non-skin colors, based on samples from a variety of ages and ethnicities.
        \\
        Sepsis (827)
        & 110,341 & 3 & 2 & 92.6 & Admissions of hospitalized subjects between 2011 and 2012 in Norway who were diagnosed with infections, systemic inflammatory response syndrome, sepsis by causative microbes, or septic shock. Target is vital status 9 days later.
        \\
        Dota2 (367)
        & 102,944 & 115 & 2 & 52.7 & Dota 2 is a computer game with two teams of 5 players. At the start of the game each player chooses a unique hero with different strengths and weaknesses. Each row represents a single game. Heroes chosen by one team are given value 1, those chosen by the other team are given value -1, and the remainder have value 0. Target is the result of the match.
        \\
        \makecell[tl]{Hospitalization\\(296)} & 101,766 & 47 & 3 & 53.9 & Ten years (1999--2008) of clinical care at 130 US hospitals. Each row represents a patient diagnosed with diabetes, who was hospitalized and discharged. The goal is to determine the early readmission of the patient within 30 days of discharge.
        \\
        Shuttle (148)
        & 58,000 & 7 & 7 & 78.6 & This data set was generated originally to extract comprehensible rules for determining the conditions under which an autolanding would be preferable to manual control of a spacecraft. The task is to decide what type of control to use. \\
        \bottomrule
    \end{tabular*}
\end{table}
Since our work only covers binary classification, datasets that had more than one response class were dichotomized by merging all classes except the largest. The observations were split into ``train", ``test", and ``calibration" sets in a ratio of 70/10/20, and a random forest with 101 trees was fitted to the training set using the \texttt{RandomForestClassifier} from \texttt{scikit-learn} \citep{sklearn}, with default values for all hyperparameters other than the number of trees. Stopping strategies for ensembles of 101 base models were computed using the minimax, minimean, and minimixed approaches described above. Allowable disagreement rates ranged from $10^{-6}$ to $10^{-1}$ in logarithmic increments (and also included $0$). The minimean and minimixed strategies require as input an estimated or assumed distribution of the number of positive base models in the ensemble; we tested them both on the empirical distribution of the calibration set and on a uniform (flat) distribution. Note that while we expect (and observe) worse performance from the uncalibrated uniform distribution, it has the practical advantage of allowing the stopping strategy to be precomputed; see Section~\ref{section:Precomputation}.

\begin{figure}
    \includegraphics[width=0.95\linewidth]{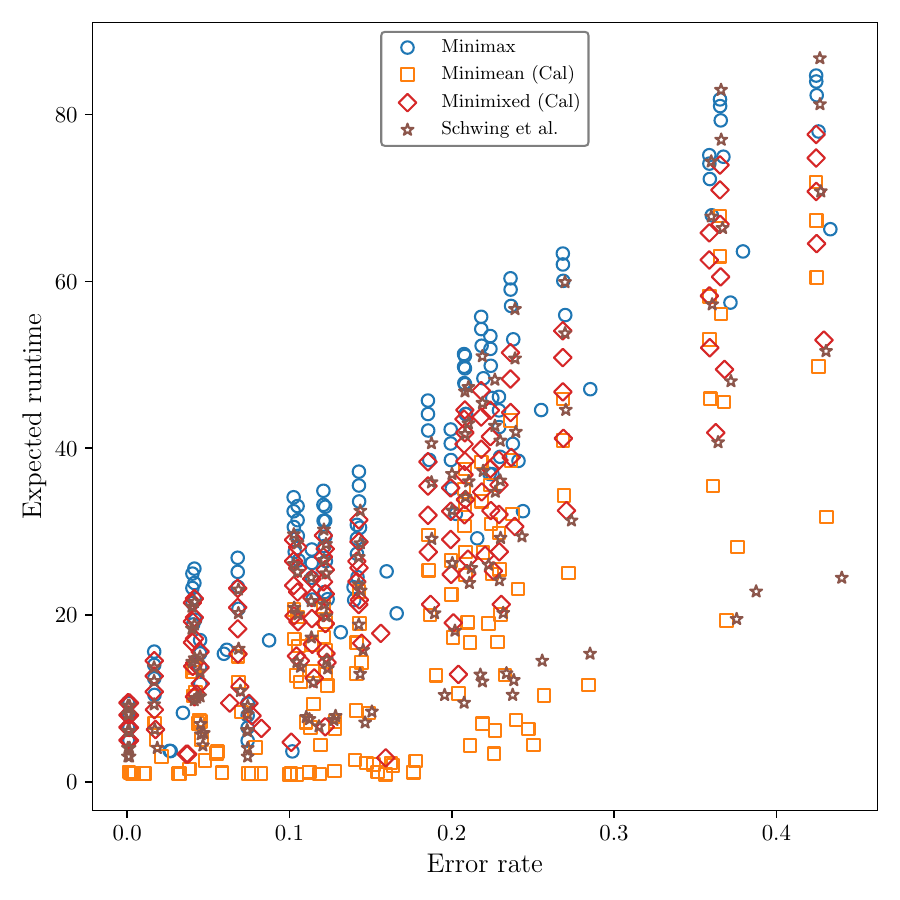}
    \caption{Time-accuracy plot comparing four early-stopping methods applied to random forests of 101 trees on 27 real-world datasets (lower is better).
    The methods are listed in Table~\ref{tab:approaches}. 
    Each point corresponds to a single method on a single dataset with a particular choice of ADR in the set $\{10^{-3}, 10^{-2.5}, 10^{-2}, 10^{-1.5}, 10^{-1}\}$.
    The X axis is the error rate $\in [0,1]$ of the resulting model and the Y axis is the expected number of trees executed before stopping ($\in [0,101]$).}
    \label{fig:er-rt-performance}
\end{figure}

The ``base" error rate is the error rate of the full ensemble on the test set; for the purposes of this experiment, the result of the full ensemble was computed by majority vote rather than by \texttt{scikit-learn}'s default behavior, which uses probability-averaging. 
The test set was also used to estimate the \textit{conditional} distribution of number of positive base models (trees) for each true response class. The expected runtime (as measured in number of base models executed) and disagreement rate were computed analytically for each possible number of positive base models; weighted by the conditional distributions, this gave us an estimated expected runtime, disagreement rate, and error rate for each stopping strategy. Note that these estimated conditional distributions were used only for estimating these metrics after the stopping strategies were computed, and not for computing the stopping strategies themselves.
This entire process, from train-test split to computation of metrics of interest, was repeated 30 times and averaged to reduce variance. The results are presented in Figure~\ref{fig:empirical-performance}.
Figure~\ref{fig:er-rt-performance} shows the tradeoff of runtime versus error rate on all the datasets considered thus far and 19 additional datasets (see supplementary). The method of \cite{SchwingZachZhengPollefeys2011} based on binomial confidence bands is included in both figures as a baseline for comparison.

\begin{figure}
    \includegraphics[width=\linewidth]{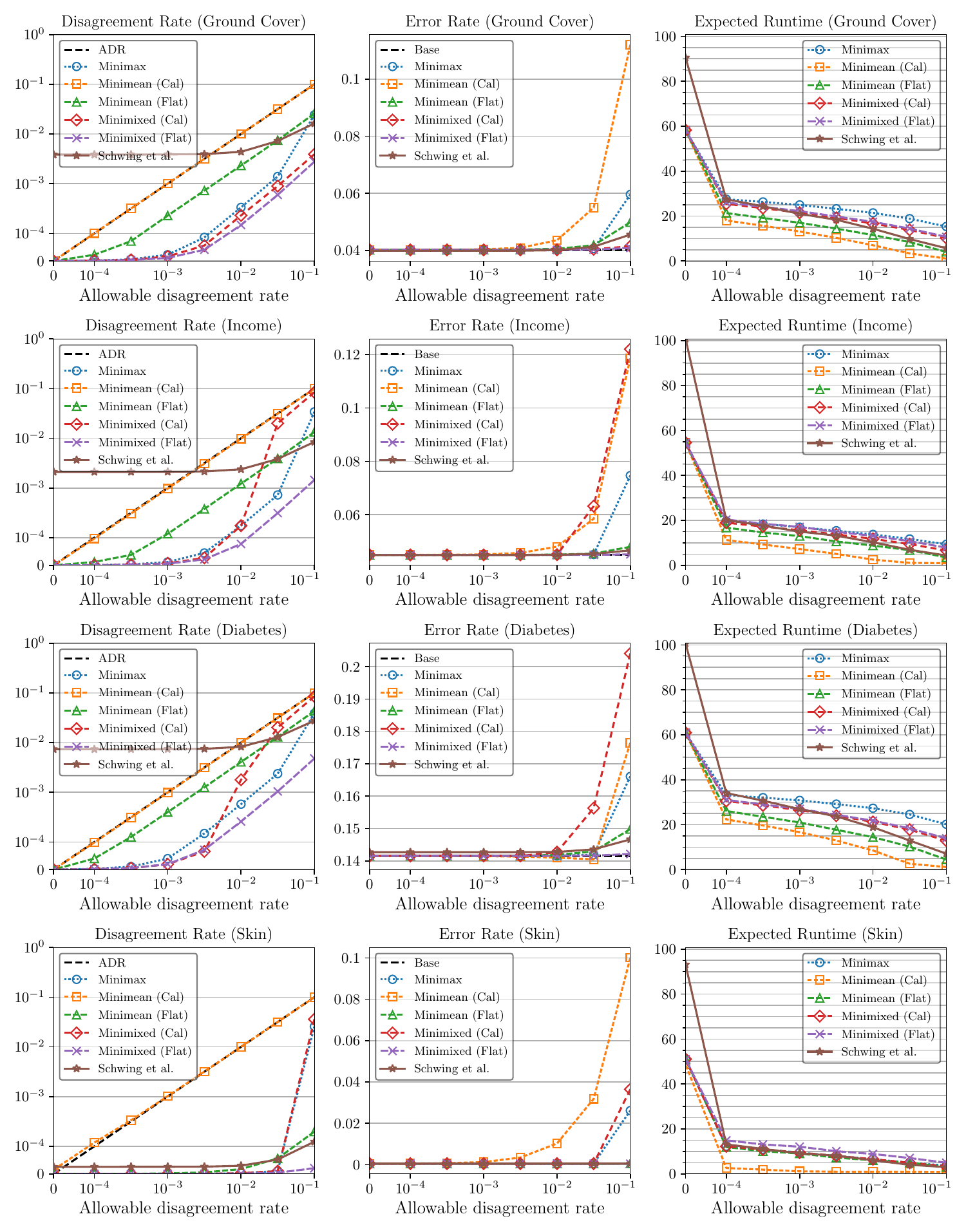}

    \caption{Performance of six early-stopping approaches for random forests of 101 trees on real-world datasets. Each row is a dataset, and each column is a metric. ``ADR" is allowable disagreement rate. ``Expected runtime" means expected number of trees executed before stopping. The approaches are listed in Table~\ref{tab:approaches}. Page 1 of 2.}
    \label{fig:empirical-performance}
\end{figure}

\begin{figure}[p]
    \vspace{-3em}

    \ContinuedFloat
    
    \includegraphics[width=\linewidth]{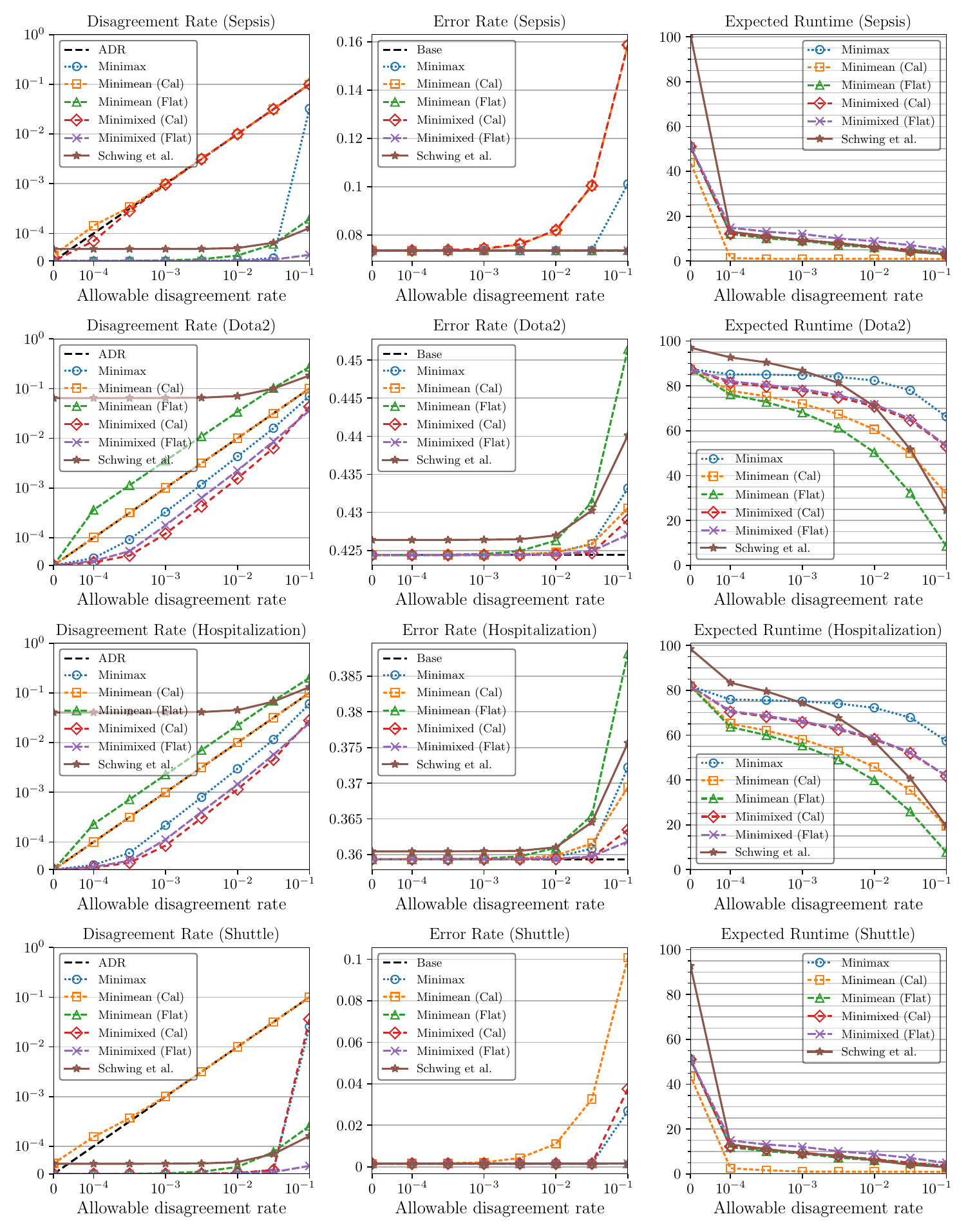}
    
    \caption{Performance of six early-stopping approaches for random forests of 101 trees on real-world datasets. Each row is a dataset, and each column is a metric. ``ADR" is allowable disagreement rate. ``Expected runtime" means expected number of trees executed before stopping. The approaches are listed in Table~\ref{tab:approaches}. Page 2 of 2.}
\end{figure}

We can see in Figure~\ref{fig:empirical-performance} that if an ADR of $10^{-3}$ or less is used, significant time can be saved with negligible excess error using any one of the approaches. With regard to runtime, we see greater savings for the minimean approach than the minimax approach, as we would expect since the minimax approach is conservative by definition. The runtime of the calibrated minimixed approach falls between the minimax and the calibrated minimean approach, and likewise for the uncalibrated approaches, as we would expect. A stopping strategy with an ADR of $10^{-3}$ will often reduce runtime by $70\%$ to $90\%$ with the minimax approach, and between $80\%$ to $99\%$ with a minimean approach using a calibration set. The exceptions were the ``Dota2" and ``Hospitalization" datasets, where time savings were smaller.

\begin{figure}[p]
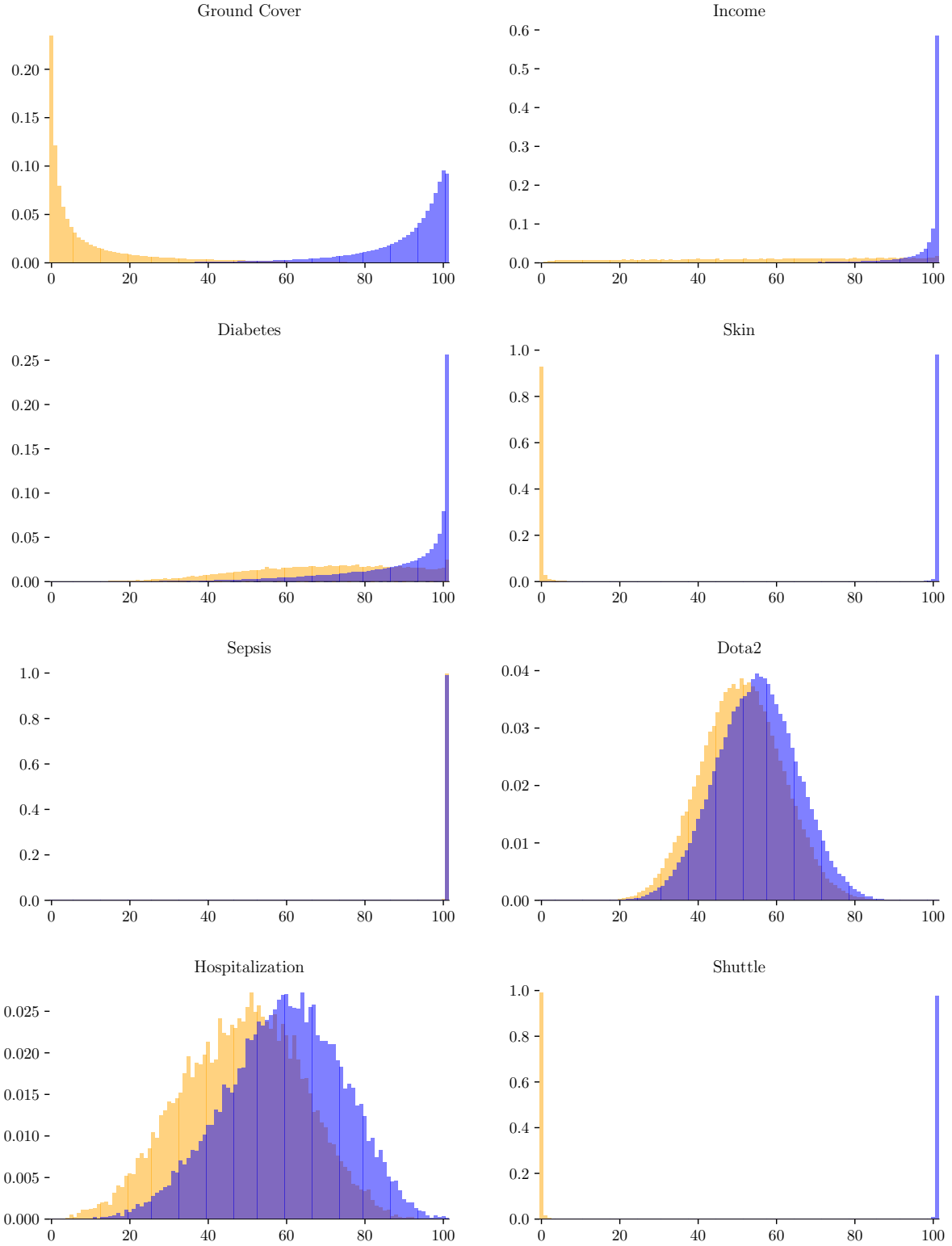

%    \resettabtoks
    \loop\ifnum\thej<4\relax
        \stepcounter{j}
        \addtabtoks{
            \includegraphics[width=0.52\linewidth]{smopdis_nonbenchmark/\thei_0.pdf} &
            \includegraphics[width=0.52\linewidth]{smopdis_nonbenchmark/\thei_1.pdf} \stepcounter{i} \\
        }
    \repeat

    \begin{tabular}{c@{\hskip -12pt} c@{\hskip -12pt} c}
        \printtabtoks
    \end{tabular}
    
    \caption{Empirical distributions of number of trees returning positive for a given observation in eight example datasets, out of a forest of 101 trees, averaged across 30 train-test splits and hundreds to thousands of observations. Orange bars represent observations whose true class is negative, and blue bars those whose true class is positive; the purple areas are overlapping bars.}
    \label{fig:smopdises}
\end{figure}

How can early stopping cut out so many base models without appreciably affecting the ensemble's output? And why does this fail on certain datasets? The answer is that, empirically for most forest-input pairs, the vast majority of trees vote the same way; this phenomenon is illustrated in Figure~\ref{fig:smopdises}. However, when the problem is particularly difficult, the distribution of tree votes is much more centralized, increasing the base ensemble's error rate \textit{and} making a consensus less likely early in the test. This is neatly illustrated by the datasets with poor runtime, ``Dota2" and ``Hospitalization": Figure~\ref{fig:empirical-performance} shows that the base error rates of these datasets were much higher than the others', and Figure~\ref{fig:smopdises} shows that their vote distributions were much closer to unimodal than the others'.

\setcounter{i}{0}
\setcounter{j}{0}

This observation also explains why the uncalibrated (flat) minimean approach generally falls between the minimax approach and the well-calibrated minimean approach in runtime. On most datasets, most ensembles are either overwhelmingly positive or overwhelmingly negative. The well-calibrated minimean approach, constructed with this information in mind, can therefore stop after only a few base models have been executed, since it is quickly obvious which of these two clusters it finds itself in. Conversely, the uncalibrated minimean approach presumes a flat distribution, so early results are taken as less indicative of the whole ensemble, making the stopping strategy more conservative, but still less so than the minimax, which by construction is \textit{maximally} conservative. This is also why the uncalibrated minimean approach is conservative on some datasets but permissive on others: when the true distribution of $n$ has a single central cluster, as in the ``Dota2" and ``Hospitalization" datasets, then the previous logic is reversed, and the assumption of a flat distribution leads to a stopping strategy that treats each vote as \textit{more} predictive of other votes than it actually is.

Taken together, these results suggest that a minimean approach using a held-out calibration set with $\alpha = 10^{-3}$ should produce good results for many datasets. Table~\ref{tab:selected-ss-metrics} summarizes this approach's performance on the datasets we tested. Expected runtime is expressed as a proportion of $N=101$, though this proportion tends to shrink as the size of the ensemble grows. ``Base error rate" is the misclassification rate of the entire ensemble without early stopping.

\begin{table}
    %\centering
    \vspace{0.75ex}
    %\vspace{-0.75ex}
    \caption{Summary of performance metrics for a minimean strategy using a calibration set.\\ The allowable disagreement rate was set to $\alpha = 10^{-3}$.}
    \label{tab:selected-ss-metrics}
    \begin{tabular}{lcccccccc}
        %\toprule
        & \rotatebox{90}{\footnotesize Ground Cover}   & \rotatebox{90}{Income}   & \rotatebox{90}{\footnotesize Diabetes}   & \rotatebox{90}{\footnotesize Skin}   & \rotatebox{90}{\footnotesize Sepsis}   & \rotatebox{90}{\footnotesize Dota2}   & \rotatebox{90}{\footnotesize Hospitalization}   & \rotatebox{90}{\footnotesize Shuttle}   \\
        \midrule
         {\footnotesize \bf Disagreement Rate} & 0.1\%          & 0.1\%    & 0.1\%      & 0.1\%  & 0.1\%    & 0.1\%   & 0.1\%             & 0.1\%     \\
         {\footnotesize \bf Expected Runtime}  & 13.02\%         & 7.17\%    & 16.50\%     & 1.15\%  & 0.99\%    & 71.32\%  & 57.58\%            & 1.03\%     \\
         {\footnotesize \bf Base Error Rate}   & 4.05\%          & 4.52\%    & 14.17\%     & 0.04\%  & 7.39\%    & 42.43\%  & 35.90\%            & 0.15\%     \\
         {\footnotesize \bf Error Rate}        & 4.07\%          & 4.54\%    & 14.16\%     & 0.12\%  & 7.46\%    & 42.43\%  & 35.90\%            & 0.21\%     \\
        \bottomrule
    \end{tabular}
    
\end{table}

\section{Discussion} \label{section:Discussion}

In this paper, after formally defining the notion of a ``stopping strategy" for binary classification ensembles, we proposed several methods to compute stopping strategies that minimize the number of base models executed while controlling the rate of disagreement with the complete ensemble. In experiments on random forests, we saw that for most datasets these methods can cut prediction runtime by 80\% or more with minimal loss of accuracy. The methods in this paper are equally applicable to ensembles of other base models; for example, ensemble methods are currently gaining popularity for test-time scaling of large language models \citep{WangEtAl2023}, including majority-voting ensembles \citep{AggarwalEtal2023,TradChehab2025}. Beyond the obvious applications for saving time and energy, this also allows the user to increase accuracy by training a larger ensemble to begin with and trusting the early-stopping strategy to only use as much of it as is necessary for each test. However, our method is limited in its applicability: it cannot be directly applied to multi-class classification problems, let alone regression problems, nor to ensembles in which the base models have a meaningful ordering or different weights.

Strictly speaking, the work in this paper is only relevant to ensembles that aggregate their base models by majority vote. The random forest as originally defined by Breiman is such an ensemble \citep{RandomForests}.
While many software packages follow his example, such as Apache Spark and the R packages \texttt{randomForest} and \texttt{ranger} \citep{LiawWiener2002, rangerR, ApacheSpark}, other implementations aggregate their results by averaging the probability estimates from each tree, so our methods cannot be directly applied to them. In particular, the Python packages scikit-learn and CuML both use this probability-averaging\footnote{For this reason, when we used scikit-learn forests for our experiments in Section~\ref{section:EmpiricalResults}, we computed the overall forest result using our own code rather than the method provided by scikit-learn itself.}, as does the cross-platform library WEKA \citep{sklearn, WEKA}. However, the performance of the two approaches is almost indistinguishable on all the datasets we have tested.
See Section~\suppref{section:BinaryVoteModel}.

As we saw in Section~\ref{section:RelatedWork}, other techniques have already been developed for speeding up ensemble classifiers, and particularly for accelerating random forests. Our method is mutually exclusive with those that involve deliberate reordering of the base models, but not with other common techniques such as pruning. However, we do not expect the gains from sequential testing to be as dramatic for a pruned ensemble, since---as discussed in Section~\ref{section:EmpiricalResults}---the current results are likely due to the fact that random forests include many trees that tend to vote the same way, a quality which pruning specifically aims to reduce. On the other hand, as discussed in Section~\suppref{section:BinomialVsHypergeometric}, most similar approaches are theoretically incompatible with pruning, whereas ours is not.

Each of the limitations mentioned presents a potential avenue for future research.
Some of the most promising extensions are to multi-class classification and weighted ensembles, as the analysis in Section~\ref{section:Theory} seems like it could be extended to either one. Another avenue would be to improve the solver to efficiently find stopping strategies for very large ensembles, on the order of $N = 1000$ or more.

\subsection*{Acknowledgments}
\addcontentsline{toc}{subsection}{Acknowledgments}

\if1\anon
We thank Yohai Bar Sinai, Saharon Rosset and David Steinberg for their helpful suggestions.
AM is supported in part by ISF Grant No. 1662/22 and NSF-BSF Grant No. 2022778.
\fi

Large Language Models (GPT-3.5, -4, o1, -4.5, -4.1, o3, -5, -5.1--5.4; all Claude models up to Opus 4.7; Gemini 2.0 Flash, 2.0 Pro, 2.5 Flash, 2.5 Pro, 2.5 Flash Lite, 3 Pro, 3 Deep Think, 3 Flash, 3.1 Pro)  were used for copyediting this manuscript, code completion, literature search and keyword selection.

\section{Reproducibility and Code Availability} \label{sec:reproducibility}

All code used in this paper may be found at the following GitHub repository:\\
\if1\anon \url{https://github.com/joseph-e-k/random-forest-early-stopping} \fi
\if0\anon \textbf{[URL omitted for anonymization]} \fi
\\
\noindent In particular, the repository contains a script which reproduces all figures and tables in the paper, except Table~\ref{tab:symbols} (the table of symbols).

\bibliography{references}
\addcontentsline{toc}{section}{References}

\end{document}

\def\spacingset#1{\renewcommand{\baselinestretch}%
{#1}\small\normalsize} \spacingset{1}

%%%%%%%%%%%%%%%%%%%%%%%%%%%%%%%%%%%%%%%%%%%%%%%%%%%%%%%%%%%%%%%%%%%%%%%%%%%%%%

\if1\anon
{
  \title{\bf Supplementary Material: Optimal Sequential Testing for Binary Ensemble Classifiers}
  \author{Joseph Kalman \href{mailto:joezkal@gmail.com}{\small \texttt{joezkal@gmail.com}}\thanks{
    We thank Yohai Bar Sinai, Saharon Rosset and David Steinberg for their helpful suggestions.
    AM is supported in part by ISF Grant No. 1662/22 and NSF-BSF Grant No. 2022778.
    Large Language Models (GPT-3.5, -4, o1, -4.5, -4.1, o3, -5, -5.1--5.4; all Claude models up to Opus 4.7; Gemini 2.0 Flash, 2.0 Pro, 2.5 Flash, 2.5 Pro, 2.5 Flash Lite, 3 Pro, 3 Deep Think, 3 Flash, 3.1 Pro)  were used for copyediting this manuscript, code completion, literature search and keyword selection.}\hspace{.2cm}\\
    Department of Statistics and Operations Research, Tel Aviv University \\
    and \\
    Amit Moscovich \href{mailto:mosco@tauex.tau.ac.il}{\small \texttt{mosco@tauex.tau.ac.il}} \\
    Department of Statistics and Operations Research, Tel Aviv University}
  \maketitle
} \fi

\if0\anon
{
  \bigskip
  \bigskip
  \bigskip
  \begin{center}
    {\LARGE\bf Supplementary Material: Optimal Sequential\\Testing for Binary Ensemble Classifiers}
\end{center}
  \medskip
} \fi

\bigskip

\vspace{8em}
\tableofcontents
\newpage

\spacingset{1.8}

\section{Proofs} \label{section:Proofs}

\subsection*{Notation}

As in Section~\ref{section:Theory} (see Table~\ref{tab:symbols}), $N$ is the number of base models in our ensemble, $n$ the number of them returning ``positive", $\theta$ the stopping strategy, $B$ the number of base models executed before stopping, and $W_{ij}$ the event that the ensemble \textit{would} reach state $(i, j)$ if unstopped. $M_i$ is a boolean random variable indicating whether the $i$-th base model returns ``positive"; note that this is nondegenerate even given $N$ and $n$ because the order of the base models is random. $A_i$ is the number of base models returning ``positive" among the first $i$, so $A_i = \sum_{i'=1}^{i} M_{i'}$; $A_i$ is also a random variable, but $A_0 = 0$ and $A_N = n$ with certainty. The random decisions of nondeterministic stopping strategies are represented by $S_{ij} \sim \Ber(\theta_{ij})$, each indicating whether the test should stop in state $(i, j)$ if that state is reached. Hence,
\env{align}{
    \reachable i j \iff &\; A_i = j \nonumber \\
    \reach{(i, j)} \iff &\; A_i = j \tand B \geq i \nonumber \\
    \stopat{(i, j)} \iff &\; A_i = j \tand B \geq i \tand S_{ij} = 1.
}

\subsection*{Proof of Lemma~\ref{claim:exists-pi-for-theta} (Necessity)}

Fix $N, n$ and $\theta \in \Theta_N$. Define $p, \pi, \bar \pi$ as in Equation~\eqref{pi-interp}. We must show that Conditions~\eqref{sky-condition-first-step}--\eqref{sky-condition-last-step} hold, which we will do one at a time. Leave the conditioning on $N, n, \theta$ implicit for what follows. For Condition~\eqref{sky-condition-first-step} we have
\env{align}{
    p_{00}
    =&\; \Prob{\reach{(0, 0)} \mid \reachable 0 0} \nonumber \\
    =&\; \Prob{A_0 = 0 \tand B \geq 0 \mid A_0 = 0} \nonumber \\
    =&\; \Prob{A_0 = 0 \mid A_0 = 0} \nonumber \\
    =&\; 1
}
as required. Condition~\eqref{sky-condition-unreachable} simply expresses the convention we have adopted for conditioning on events of probability zero, since $\Prob{\reachable i j} = 0$ wherever $j > i$. For Condition~\eqref{sky-condition-pi-plus-pi-bar} we have for each $i, j \leq i$
\env{align}{
    \pi_{ij} + \bar \pi_{ij}
    =&\; \Prob{\reach{(i, j)} \tand \stopat{(i, j)} \mid \reachable i j} \nonumber \\
        &+ \Prob{\reach{(i, j)} \tand \tnot \stopat{(i, j)} \mid \reachable i j} \nonumber \\
    =&\; \Prob{\reach{(i, j)} \mid \reachable i j} \nonumber \\
    =&\; p_{ij}
}
as required. For Condition~\eqref{sky-condition-last-step}, note that $\theta \in \Theta_N$ and so $\theta_{Nj} = 1$, or in other words $\Prob{S_{Nj} = 1} = 1$, for all $j$. For each $j$ we therefore have
\env{align}{
    \frac{\pi_{Nj}}{p_{Nj}}
    =&\; \frac{\Prob{\reach{(N, j)} \tand \stopat{(N, j)} \mid \reachable N j}}{\Prob{\reach{(N, j)} \mid \reachable N j}} \nonumber \\
    =&\; \frac{\Prob{A_N = j \tand B \geq N \tand S_{Nj} = 1 \mid A_N = j}}{\Prob{A_N = j \tand B \geq N \mid A_N = j}} \nonumber \\
    =&\; \frac{\Prob{A_N = j \tand B \geq N \mid A_N = j}}{\Prob{A_N = j \tand B \geq N \mid A_N = j}} \nonumber \\
    =&\; 1
}
as required. Condition~\eqref{sky-condition-combinatoric} is slightly more complicated. We begin by establishing
\env{align}{
    p_{i+1, j+1}
    =&\; \Prob{\reach{(i + 1, j + 1)} \mid \reachable {i+1} {j+1}} \nonumber \\
    =&\; \Prob{B \geq i + 1 \mid A_{i+1} = j + 1} \nonumber \\
    =&\; \Prob{B > i \mid A_{i+1} = j + 1}.
}
By complete expectation over values of $A_i$ this becomes
\env{align}{
    p_{i+1, j+1}
    =&\; \bigsum{j'=0}{N} \Prob{A_i = j' \mid A_{i+1} = j + 1} \Prob{B > i \mid A_i = j', A_{i+1} = j + 1}.
}
Since $A_{i+1}$ must be equal to either $A_i$ or $A_i+1$, the term $\Prob{A_i = j' \mid A_{i+1} = j + 1}$ is zero for all $j'$ other than $j' = j$ and $j' = j+1$, so this becomes
\env{align}{
    p_{i+1, j+1}
    =&\;\Prob{A_i = j \mid A_{i+1} = j + 1} \Prob{B > i \mid A_i = j, A_{i+1} = j + 1} \\
        &+ \Prob{A_i = j + 1 \mid A_{i+1} = j + 1} \Prob{B > i \mid A_i = j + 1, A_{i+1} = j + 1}. \nonumber
}
Note that the event $B > i$ depends on $A_{i+1}$ only through $A_i$, so we can drop one of the conditions in each term, giving us
\env{align}{
    p_{i+1, j+1}
    =&\; \Prob{A_i = j \mid A_{i+1} = j + 1} \Prob{B > i \mid A_i = j} \\
        &+ \Prob{A_i = j + 1 \mid A_{i+1} = j + 1} \Prob{B > i \mid A_i = j + 1}. \nonumber
}
Recall that $A_i = \sum_{i'=1}^{i} M_{i'}$. Therefore, if $A_{i+1} = j + 1$, then $A_i = j$ iff $M_{i + 1} = 1$. Therefore we can substitute the events to get
\env{align}{
    p_{i+1, j+1}
    =&\; \Prob{M_{i+1} = 1 \mid A_{i+1} = j + 1} \Prob{B > i \mid A_i = j} \\
        &+ \Prob{M_{i+1} = 0 \mid A_{i+1} = j + 1} \Prob{B > i \mid A_i = j + 1}. \nonumber
}
Note now that by simple combinatorics we have
\env{align}{
    \Prob{M_i = 1 \mid A_i = j} =&\; \Prob{M_i = 1 \mid \bigsum{i'=1}{i} M_{i'} = j} = \frac{j}{i} \nonumber \\
    \implies \Prob{M_i = 0 \mid A_i = j} =&\; 1 - \frac{j}{i} = \frac{i - j}{i}
}
and likewise if we substitute $j + 1$ for $j$. Therefore we have
\env{align}{
    p_{i+1, j+1}
    =&\; \frac{j + 1}{i + 1} \Prob{B > i \mid A_i = j} \\
        &+ \frac{i - j}{i + 1} \Prob{B > i \mid A_i = j + 1} \nonumber \\
    =&\; \frac{j + 1}{i + 1} \bar \pi_{ij} + \frac{i - j}{i + 1} \bar \pi_{i, j+1} \nonumber
}
as required. Thus all four conditions are satisfied and $(p, \pi, \bar \pi) \in \Pi$ as required. \qed

\subsection*{Proof of Lemma~\ref{claim:can-find-theta-for-pi} (Sufficiency)}

Fix $N, n$ and let $(p, \pi, \bar \pi) \in \Pi$. Define $\theta$ as follows:
\env{align}{
    \theta_{ij} := \env{cases}{
        1,& p_{ij} = 0 \\
        \frac{\pi_{ij}}{p_{ij}},& p_{ij} \neq 0.
    }
}
Clearly, this $\theta$ can be computed in $O(N^2)$ time. Moreover, by Condition~\eqref{sky-condition-pi-plus-pi-bar} for each $i, j$ we have $p_{ij} = \pi_{ij} + \bar \pi_{ij} \geq \pi_{ij}$ and thus $\theta_{ij} \in [0, 1]$. By Condition~\eqref{sky-condition-last-step} for each $j$ we have $\pi_{Nj} = p_{Nj}$ and so
\env{align}{
    \theta_{Nj} = \env{cases}{
        1,& p_{Nj} = 0 \\
        \frac{\pi_{Nj}}{p_{Nj}},& p_{Nj} \neq 0
    } ~= 1,
}
so $\theta \in \Theta_N$ as required. All that remains is to show that Equation~\eqref{pi-interp} holds. By the construction of $\theta$ we have $\pi_{ij} = p_{ij} \theta_{ij}$ for each $i, j$. Thus, if $p_{ij} = \Prob{B \geq i \mid A_i = j}$, then $\pi_{ij} = \Prob{B = i \mid A_i = j}$. The equivalent holds for $\bar \pi$. Therefore, once the first equation in \eqref{pi-interp} has been proven for a certain $i$, the other two follow immediately. We now proceed recursively, demonstrating that $p_{ij} = \Prob{B \geq i \mid A_i = j}$ for each step $i$ starting from $i = 0$. As a base case, from Condition~\eqref{sky-condition-first-step} we have for each $j$
\env{align}{
    \Prob{\stopat{(0, j)} \mid \reachable 0 j} = \Prob{B \geq 0 \mid A_0 = j} = 1 = p_{0j},
}
so the desideratum holds for each $p_{0j}$. At subsequent steps, we start by applying Condition~\eqref{sky-condition-combinatoric} to get
\env{align}{
    p_{i + 1, j + 1}
    =& \; \frac{j + 1}{i + 1} \bar \pi_{ij} + \frac{i - j}{i + 1} \bar \pi_{i, j + 1}
}
then employ the same combinatoric argument as in the proof of Lemma~\ref{claim:exists-pi-for-theta} to get
\env{align}{
    \Prob{A_i = j \mid A_{i+1} = j + 1} =&\; \frac{j + 1}{i + 1} \\
    \Prob{A_i = j + 1 \mid A_{i+1} = j + 1} =&\; 1 - \frac{j + 1}{i + 1} = \frac{i - j}{i + 1},
}
giving us
\env{align}{
    p_{i + 1, j + 1}
    =& \; \Prob{A_i = j \mid A_{i + 1} = j + 1} \bar \pi_{ij} \\
    & \; + \Prob{A_i = j + 1 \mid A_{i + 1} = j + 1} \bar \pi_{i, j + 1}. \nonumber
}
By the assumption of recursion this becomes
\env{align}{
    p_{i + 1, j + 1}
    =& \; \Prob{A_i = j \mid A_{i + 1} = j + 1} \Prob{B > i \mid A_i = j} \\
    & \; + \Prob{A_i = j + 1 \mid A_{i + 1} = j + 1} \Prob{B > i \mid A_i = j + 1}, \nonumber
}
and substituting $B > i \iff B \geq i + 1$ gives us
\env{align}{
    p_{i + 1, j + 1}
    =& \; \Prob{A_i = j \mid A_{i + 1} = j + 1} \Prob{B \geq i + 1 \mid A_i = j} \\
    & \; + \Prob{A_i = j + 1 \mid A_{i + 1} = j + 1} \Prob{B \geq i + 1 \mid A_i = j + 1}, \nonumber
}
which---again by the same argument as for the previous lemma---is simply a total probability over antecedent states of $A_{i + 1} = j + 1$, simplifying to
\env{align}{
    p_{i + 1, j + 1}
    =& \; \Prob{B \geq i + 1 \mid A_{i + 1} = j + 1} \nonumber \\
    =& \; \Prob{\reach{(i + 1, j + 1)} \mid \reachable {i+1} {j+1}}
}
as required. \qed

\subsection*{Proof of Theorem~\ref{claim:minimax-problem-pi-theta-equivalent}}

Suppose $(p^*, \pi^*, \bar \pi^*)$ is an optimal solution to Problem~$\Pi$-Minimax \eqref{problem:pi-minimax}. By Lemma~\ref{claim:can-find-theta-for-pi}, we can find some $\theta^* \in \Theta_N$ that has the same objective value in $O(N^2)$ time. Suppose by way of contradiction that this $\theta^*$ is not actually an optimal solution for Problem~$\Theta$-Minimax \eqref{problem:theta-minimax}; then there exists some other $\tilde \theta \in \Theta_N$ whose objective value is lower. But if so, then by Lemma~\ref{claim:exists-pi-for-theta}, there exists some $(\tilde p, \tilde \pi, \tilde {\bar \pi}) \in \Pi$ which has the same objective value as $\tilde \theta$, and therefore has a better objective value than $(p^*, \pi^*, \bar \pi^*)$, contradicting the optimality of $(p^*, \pi^*, \bar \pi^*)$. \qed

\section{Pseudocode} \label{section:Pseudocode}

The basic outline of our method, described in Algorithm~\ref{alg:compute-oss}, is the same regardless of whether we take a minimax, minimean, or minimixed approach: we construct a linear program, solve it, and convert the solution to a stopping strategy. This algorithm accepts the ensemble size $N$, allowable disagreement rate $\alpha$,  distribution of positive votes $\hat D$, and two binary parameters $f_r, f_d$ which specify the approach (worst-case or expectation) to use for the expected runtime and disagreement rate respectively, and returns the stopping strategy which is optimal in the desired sense.

\begin{algorithm}
    \caption{(\textsc{ComputeOSS}) Find an optimal stopping strategy.} \label{alg:compute-oss}
    \begin{algorithmic}[1]
        \Require \parbox[t]{\linewidth}{$N \in \setn,\, \alpha \in [0, 1],\, \hat D \in [0, 1]^{N + 1}, \\ f_r \in \set{\text{min}, \text{mean}},\, f_d \in \set{\text{min}, \text{mean}}$}
        \State $P \gets \Call{MakeOSSProgram}{N, \alpha, \hat D, f_r, f_d}$
        \State $p^*, \pi^*, ... \gets \Call{SolveLinearProgram}{P}$
        \State $\theta^* \gets \Call{ConvertPiToTheta}{p^*, \pi^*}$
        \State \Return $\theta^*$
    \end{algorithmic}
\end{algorithm}
The first step is setting up the linear program (\textsc{MakeOSSProgram}) as described in Algorithm~\ref{alg:construct-osp}. 
\begin{algorithm}
    \caption{(\textsc{MakeOSSProgram}) Construct a linear program representing the optimal stopping problem.} \label{alg:construct-osp}
    \begin{algorithmic}[1]
        \Require \parbox[t]{\linewidth}{$N \in \setn,\, \alpha \in [0, 1],\, \hat D \in [0, 1]^{N + 1}, \\ f_r \in \set{\text{min}, \text{mean}},\, f_d \in \set{\text{min}, \text{mean}}$}
        
        \State $P \gets \Call{Linearprogram}$
        \State $p, \pi, \bar \pi \gets \Call{SetUpDecisionVariables}{P, N}$ \Comment{See Algorithm~\ref{alg:set-up-decision-variables}}
        
        \For{$0 \leq n \leq N,~ 0 \leq i \leq N,~ 0 \leq j \leq N$}
            \State $a_{nij} \gets \Call{HGProb}{N, n, i, j}$ \Comment{$\Prob{\reachable i j \mid N, n}$ per \eqref{formula:prob-reachable}}
            \State $s_{nij} \gets \pi_{ij} a_{nij}$\Comment{$\Prob{\stopat{(i, j)} \mid N, n, \theta}$ per \eqref{formula:prob-stop-at}}
            \State $d_{nij} \gets \brac{\brac{j > i / 2} \neq \brac{n > N / 2}}$ \Comment{Disagreement indicator}
        \EndFor

        \For{$0 \leq n \leq N$}
            \State $E_n \gets \sum_{i=0}^{N} i \cdot \sum_{j=0}^{N} s_{nij}$ \Comment{$\E[B \mid N, n, \theta]$ per \eqref{formula:expected-B}}
            \State $q_n \gets \sum_{i=0}^{N} \sum_{j=0}^{N} d_{nij} s_{nij}$ \Comment{$\Prob{\text{disagree} \mid N, n, \theta}$ per \eqref{formula:prob-disagree}}
        \EndFor

        \If {$f_d = \text{min}$}
            \For{$0 \leq n \leq N$}
                \State $\Call{AddConstraint}{P,\, q_n \leq \alpha}$ \Comment{$\Prob{\text{disagree} \mid N, n, \theta} \leq \alpha$}
            \EndFor
        \Else
            \State $q' \gets \sum_{n=0}^N \hat D_n q_n$ \Comment{$\E_n \brac{\Prob{\text{disagree} \mid N, n, \theta}}$}
            \State $\Call{AddConstraint}{P,\, q' \leq \alpha}$
        \EndIf

        \If {$f_r = \text{min}$}
            \State $E' \gets \Call{DecisionVariable}{P,\, \setr}$ \Comment{$\Max{n} \E[B \mid N, n, \theta]$}
            \For{$0 \leq n \leq N$}
                \State $\Call{AddConstraint}{P,\, E' \geq E_n}$
            \EndFor
        \Else
            \State $E' \gets \sum_{n=0}^N \hat D_n E_n$ \Comment{$\E_n \brac{\E[B \mid N, n, \theta]}$}
        \EndIf
        
        \State $\Call{SetMinimizationObjective}{P,\, E'}$
        
        \State \Return P
    \end{algorithmic}
\end{algorithm}
The algorithm calls a subroutine $\textsc{HGProb}(m, m', k, k')$ which computes the hypergeometric probability mass, i.e., the probability of drawing exactly $k'$ marked objects when drawing $k$ objects at random out of a pool of $m$ of which $m'$ are marked:
\env{align}{
    \textsc{HGProb}(m, m', k, k') =&\; \frac{\binom{m'}{k'} \binom{m-m'}{k-k'}}{\binom{m}{k}}.
}
It also uses a subroutine \textsc{SetUpDecisionVariables}, which creates the matrices of decision variables $p, \pi, \bar \pi$ and adds the constraints that force them to be in $\Pi$ (defined in Equation~\eqref{formula:big-pi}); this auxiliary function is given as Algorithm~\ref{alg:set-up-decision-variables}.

\begin{algorithm}
    \caption{(\textsc{SetUpDecisionVariables}) Create decision variables and constrain them to be in $\Pi$, the set of valid values for $(p, \pi, \bar \pi)$.}\label{alg:set-up-decision-variables}
    \begin{algorithmic}[1]
        \Require $P \in \mathsf{LinearPrograms},\, N \in \setn$
        \State $p \gets \Call{DecisionVariable}{P,\, [0, 1]^{(N + 1) \times (N + 1)}}$
        \State $\pi \gets \Call{DecisionVariable}{P,\, [0, 1]^{(N + 1) \times (N + 1)}}$
        \State $\bar \pi \gets \Call{DecisionVariable}{P,\, [0, 1]^{(N + 1) \times (N + 1)}}$

        \For{$0 \leq i \leq N,~ 0 \leq j \leq N$}
            \State $\Call{AddConstraint}{P,\, p_{0j} = 1}$ \Comment{Condition~\eqref{sky-condition-first-step}}
            \State $\Call{AddConstraint}{P,\, \pi_{ij} + \bar \pi_{ij} = p_{ij}}$ \Comment{Condition~\eqref{sky-condition-pi-plus-pi-bar}}
            \If{$j < N \tand i < N$}
                \State $\Call{AddConstraint}{P,\, p_{i + 1, j + 1} = \frac{i - j}{i + 1} \bar \pi_{i, j + 1} + \frac{j + 1}{i + 1} \bar \pi_{ij}}$ \Comment{Condition~\eqref{sky-condition-combinatoric}}
            \EndIf
            \State $\Call{AddConstraint}{P,\, \pi_{Nj} = p_{Nj}}$ \Comment{Condition~\eqref{sky-condition-last-step}}
        \EndFor
        \State \Return $p, \pi, \bar \pi$
    \end{algorithmic}
\end{algorithm}

Once the linear program is constructed, we solve it. Linear programs are well-studied and many algorithms for solving them exist; see \cite{HillierLieberman2024}. We take no stance on which LP algorithm is most appropriate for our case except to note that some of the coefficients can become much smaller than others for large ensembles, potentially leading to loss of numeric precision. After solving the linear program we find the stopping strategy $\theta^*$ that induces the optimal conditional probabilities $p^*, \pi^*$. This procedure is presented as Algorithm~\ref{alg:convert-pi-to-theta}. The proof that the resulting $\theta^*$ does in fact induce $p^*, \pi^*$ as conditional probabilities can be found as part of the proof of Lemma~\ref{claim:can-find-theta-for-pi}.

\begin{algorithm}
    \caption{(\textsc{ConvertPiToTheta}) Compute a stopping strategy which induces the conditional probabilities $p, \pi$.}\label{alg:convert-pi-to-theta}
    \begin{algorithmic}[1]
        \Require $p \in [0, 1]^{(N + 1) \times (N + 1)}, \pi \in [0, 1]^{(N + 1) \times (N + 1)}$
        
        \State $\theta \gets \Call{Empty}{[0, 1]^{(N + 1) \times (N + 1)}}$

        \For{$0 \leq i \leq N,~ 0 \leq j \leq N$}
            \If {$p_{ij} = 0$}
                \State $\theta_{ij} \gets 1$
            \Else
                \State $\theta_{ij} \gets \frac{\pi_{ij}}{p_{ij}}$
            \EndIf
        \EndFor
        
        \State \Return $\theta$
    \end{algorithmic}
\end{algorithm}

These problems can be streamlined somewhat by dropping the decision variables for unreachable states (that is, states $(i, j)$ where $j > i$) and by noting that for the purposes of constraining worst-case disagreement rate, only the values $\lfloor{N / 2}\rfloor, \lfloor{N / 2}\rfloor + 1$ need to be considered for $n$, due to the following lemma (lemma~\ref{claim:need-only-consider-worst-case-n}).

\newpage

\begin{lemma}
\label{claim:need-only-consider-worst-case-n}
    For any stopping strategy $\theta \in \Theta_N $, we have \\ for all $n > N / 2$, $\Prob{\text{disagree} \mid N, n, \theta} \leq \Prob{\text{disagree} \mid N, n = \lfloor{N / 2}\rfloor + 1, \theta}$; and \\ for all $n \leq N / 2$, $\Prob{\text{disagree} \mid N, n, \theta} \leq \Prob{\text{disagree} \mid N, n = \lfloor{N / 2}\rfloor, \theta}$.
\end{lemma}

\subsection*{Proof of Lemma~\ref{claim:need-only-consider-worst-case-n}}

Fix $N, \theta$. For each value of $n$, a ``disagreement state" is a state where the early-stopping conclusion would differ from that of the complete ensemble. The event of disagreement is just the union, over all disagreement states, of reaching and stopping at that state. Since stopping in one state is mutually exclusive with stopping in another state, the probability of disagreement is simply the sum over all disagreement states of the probability of reaching and stopping in that state. In other words,
\env{align}{
    \Prob{\text{disagree}}
    =&\; \bigsum{i=0}{N} \bigsum{j=0}{i} \ind{\ind{j > i / 2} \neq \ind{n > N / 2}} \Prob{\stopat (i, j)}.
}
Assume that $n > N / 2$. We will show, for \textit{each} disagreement state, that the probability of stopping in that state is maximized by $n = \floor{N / 2} + 1$ among all values of $n$ satisfying our assumption.

Let $(i, j)$ be a disagreement state, that is, let $j \leq i /2$. Note that we can only stop at state $(i, j)$ if precisely $j$ of the first $i$ base models are positive, an event we call $W_{ij}$. Therefore the probability of stopping in that state, as a function of $n$, is
\env{align}{
    \Prob{\stopat (i, j) \mid N, n, \theta}
    =&\; \Prob{W_{ij} \mid N, n, \theta} \Prob{\stopat (i, j) \mid N, n, \theta, W_{ij}}.
}
We can simplify the conditions by noting that the event $W_{ij}$ is independent of $\theta$, and that---given $W_{ij}$---the probability of stopping at $(i, j)$ is independent of $n$. To prove the latter, consider that---given $W_{ij}$---each possible ordering of $j$ positive base models and $(i - j)$ negative ones among the first $i$ base models is equally likely, regardless of $n$. Given some ordering of these first $i$ base models, the probability of reaching and stopping at $(i, j)$ depends only on the results of the stopping decisions $S_{i'j'}$ for $i' \leq i$ and $j'$ determined by that ordering. The random variables $S_{i'j'}$, in turn, have distributions that depend only on $\theta$ and are otherwise independent of everything. Thus the probability of stopping at $(i, j)$ does not depend on $n$,
\env{align}{
    \Prob{\stopat (i, j) \mid N, n, \theta}
    =&\; \Prob{W_{ij} \mid N, n} \Prob{\stopat (i, j) \mid \theta, W_{ij}}.
    %\\\propto&\; \Prob{W_{ij} \mid N, n}.
}
The probability of $W_{ij}$ is given by the probability mass function of the hypergeometric distribution. Choosing $n$ that maximizes $\Prob{W_{ij} \mid N, n}$ is thus equivalent to finding a maximum likelihood estimator for $n$ given the observation that $j$ of the first $i$ base models were positive. The likelihood function of the number of ``positive" objects in the hypergeometric distribution is known to be unimodal with a peak where the proportion of positives in the population ($n / N$) matches that in the sample ($j / i$). Since we have assumed that $n > N / 2$, and conversely that $j \leq i / 2$, the closest we can bring $n / N$ to $j / i$ is by setting $n = \floor{N / 2} + 1$.
This argument applies to each disagreement state separately. Summing over all disagreement states, we see that the probability of disagreement is maximized by $n = \floor{N / 2} + 1$ among all values of $n > N / 2$. A symmetrical argument shows that the probability of disagreement is maximized by $n = \floor{N / 2}$ among all values of $n \leq N / 2$.

Therefore, if we are given $N$ and want to constrain $\theta$ to keep the disagreement rate below $\alpha$ for all values of $n$, it is sufficient to keep the disagreement rate below $\alpha$ for $n \in \set{\floor{N / 2}, \floor{N / 2} + 1}$.

\qed

\section{LP Runtime Compared to QCQP Baseline} \label{section:SolutionTimes}

Problems $\Theta$-Minimax, $\Theta$-Minimean, and $\Theta$-Minimixed (Problems \eqref{problem:theta-minimax}, \eqref{problem:theta-minimean}, and \eqref{problem:theta-minimixed} respectively) are not linear programs, nor even quadratic programs. However, they can easily be reformulated as quadratically-constrained quadratic programs (QCQPs) with $O(N^3)$ decision variables. 
However, our experiments found that solving these QCQPs was extremely slow even for small values of $N$. In this Section, we discuss these experiments in more detail.
To use the minimax problems as an example, recall Problem~$\Theta$-Minimax \eqref{problem:theta-minimax}:
\env{align}{
    \minimize_{\theta \in \Theta_N}
    \ \max_n
    \ \E[B \mid N, n, \theta]
    \qquad
    \text{s.t.}
    \qquad
    \max_n
    \Pr(\text{disagree} \mid N, n, \theta) \leq \alpha. \nonumber
}
\noindent where
\env{align}{
    \E\brac{B \mid N, n, \theta}
    &=\; \bigsum{i=0}{N} i \cdot \bigsum{j=0}{i} \Prob{\stopat{(i, j)} \mid N, n, \theta}, \nonumber \\
    \Prob{\text{disagree} \mid N, n, \theta}
    &= \bigsum{i=0}{N} \bigsum{j=0}{i} \Prob{\stopat{(i, j)} \mid N, n, \theta} d_{ij}(N, n) \nonumber.
}

\noindent
Both the objective and the constraint of the problem are linear combinations of the conditional stopping probabilities $\Prob{\stopat{(i, j)} \mid N, n, \theta}$ for various values of $(i, j)$. However, $\Prob{\stopat{(i, j)} \mid N, n, \theta}$ must be computed recursively: to reach and stop in a state we must first reach and \textit{not} stop in one of its two antecedent states, and so on back to the states of the first step $i = 0$. Since the probability of stopping in a state $(i, j)$ if reached is $\theta_{ij}$ by definition, computing $\Prob{\stopat{(i, j)} \mid N, n, \theta}$ requires us to multiply together $i$ different $\theta$ variables, yielding an expression that is not a linear or quadratic combination of decision variables.

\begin{figure}
    \includegraphics[width=0.95\linewidth]{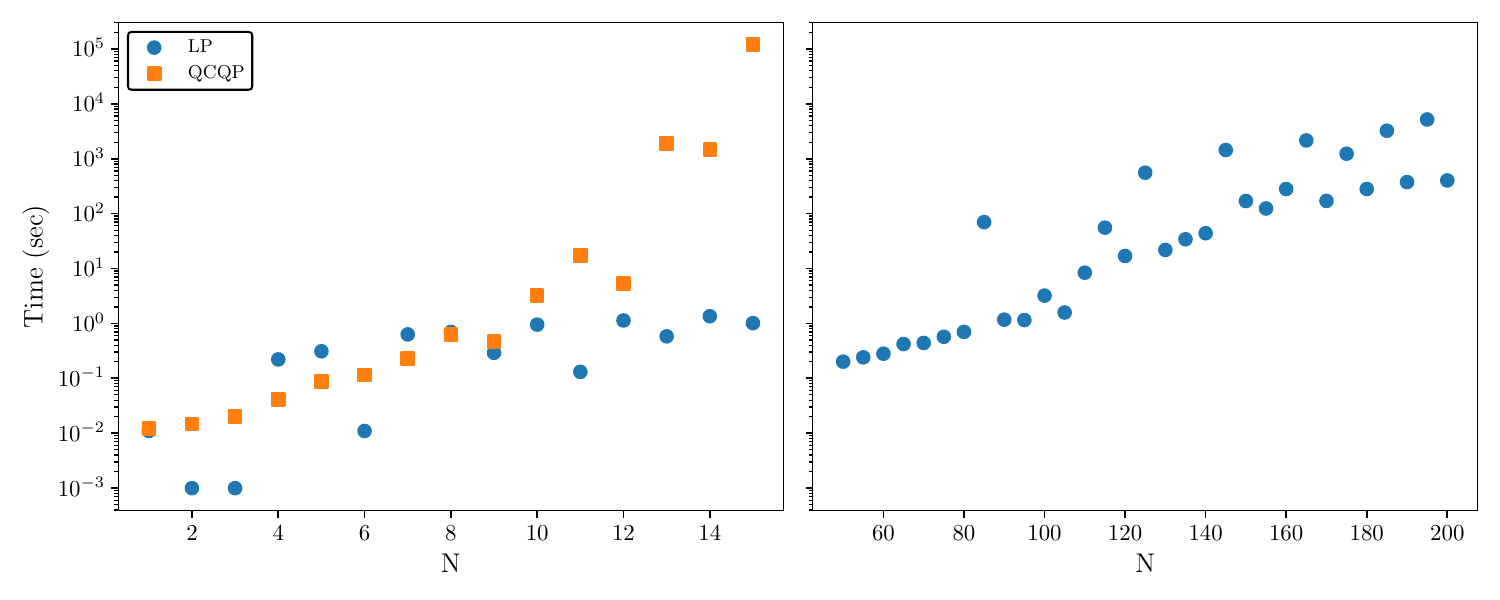}
    \caption{Runtime to find a minimax-optimal stopping strategy for an ensemble of size $N$ with $\alpha = 0.01$ using a quadratically-constrained quadratic program (QCQP) compared to the time using a linear program (LP), on a log scale. Each point is the minimum of 3 runs.}
    \label{fig:timing}
\end{figure}

We can make the problem a QCQP by adding another set of decision variables to represent the conditional stopping probabilities $\Prob{\stopat{(i, j)} \mid N, n, \theta}$ directly, and then adding constraints to ensure that these decision variables have the required relationship with the $\theta$ variables. This brings the number of decision variables to $O(N^3)$, since we would need one variable for each value of the triplet $(i, j, n)$. To check the solvability of this QCQP, we timed solutions to the problem for various values of $N$ using the Gurobi optimization suite, version 12.0.1, on an AMD EPYC 9654 processor. The results up to $N = 15$ are given in Figure~\ref{fig:timing}. Note the logarithmic scale of the $y$-axis; the rate of growth is clearly exponential or superexponential in $N$, and the solution time is already about 33 hours for $N = 15$, compared to less than a second using a linear program. The solution times for the linear programs were also roughly $O(\exp(cN))$, but with a much smaller $c$; these results, up to $N = 200$, are also in Figure~\ref{fig:timing}.
We were pleased to see---as a sanity check---that the stopping strategies returned by the quadratic programs and by the linear programs were identical.

\pagebreak
\section{Binary Vote Model} \label{section:BinaryVoteModel}

The work in this paper assumes an ensemble where each base model's vote is binary (positive or negative), and the overall ensemble result is the majority vote, like Breiman's original random forest. However, some software implementations of the random forest, such as scikit-learn for Python or the cross-platform library WEKA, instead have each base model return a probability, and take the average of those probabilities for the final prediction (rounding to zero or one if a binary prediction is required) \citep{sklearn, WEKA}. In practice it is straightforward to work around this problem by writing one's own code to round and count votes rather than averaging them, but naturally we would be hesitant to do this if it had a nonnegligible chance of changing the ensemble's overall behavior. This raises the question: do the probabilistic votes allowed by scikit-learn and similar implementations actually change the behavior of the ensemble? Our experiments, detailed below, answer this question in the negative: in practice, these libraries can be treated as if they used binary voting.

We investigated this question using scikit-learn random forests grown with default hyperparameters. We looked at two metrics: the proportion of individual tree predictions which were certain (i.e., equal to exactly 0 or 1), and the proportion of overall classifications that were changed by rounding the individual tree predictions to 0 or 1 and taking a majority vote, as opposed to averaging the individual tree predictions and then selecting the highest-probability class. Table~\ref{tab:tree-certainty} shows these metrics for 30 random forests trained and tested on each of the datasets we looked at (including those in our main paper and the benchmark datasets proposed by \citet{RandomForestsBetterThanDL}; see their appendix on arXiv for details).

These results show that, on all datasets we examined, and when using default hyperparameters, the fact that scikit-learn theoretically allows probabilistic votes from individual trees has essentially no impact.
In other words, we lose almost no information by rounding these votes to zero or one. On almost all datasets this was due to the fact that individual trees almost always voted 0 or 1 anyway. The single exception is the ``Sepsis" dataset (UCIML dataset \#827), where most individual tree votes were uncertain, but nevertheless the overall behavior was unaffected by rounding because almost all votes were \textit{nearly} certain: over 90\% of individual tree votes were greater than 0.85 or less than 0.15.

\begin{table}[!t]
    \vspace{1ex}
    %\vspace{1em}
    \renewcommand{\arraystretch}{1.2}
    \caption{The impact of allowing probabilistic votes (versus rounding all votes to 0 or 1) on random forests applied to real-world datasets. ``Tree certainty" means the proportion of individual tree votes that were equal to zero or one. ``Invariance to rounding" means the proportion of whole-forest predictions that were unchanged by rounding the individual tree votes. \label{tab:tree-certainty}}
    \centering
    \begin{tabular*}{0.8\textwidth}{lcc}
        \toprule
        \bf Dataset name & \bf Tree certainty & \bf Invariance to rounding \\
        \midrule
        Ground Cover & 100.00\% & 100.00\% \\
        Income & 100.00\% & 100.00\% \\
        Diabetes & 99.01\% & 100.00\% \\
        Skin & 99.99\% & 100.00\% \\
        Sepsis & 12.77\% & 100.00\% \\
        Dota2 & 100.00\% & 100.00\% \\
        Hospitalization & 100.00\% & 100.00\% \\
        Shuttle & 99.72\% & 100.00\% \\
        credit & 100.00\% & 100.00\% \\
        california & 100.00\% & 100.00\% \\
        wine & 100.00\% & 100.00\% \\
        electricity & 100.00\% & 100.00\% \\
        covertype & 100.00\% & 100.00\% \\
        pol & 100.00\% & 100.00\% \\
        house\_16H & 100.00\% & 100.00\% \\
        kdd\_ipums\_la\_97-small & 100.00\% & 100.00\% \\
        MagicTelescope & 100.00\% & 100.00\% \\
        bank-marketing & 100.00\% & 100.00\% \\
        phoneme & 100.00\% & 100.00\% \\
        MiniBooNE & 99.76\% & 100.00\% \\
        Higgs & 100.00\% & 100.00\% \\
        eye\_movements & 100.00\% & 100.00\% \\
        jannis & 100.00\% & 100.00\% \\
        KDDCup09\_upselling & 100.00\% & 100.00\% \\
        rl & 100.00\% & 100.00\% \\
        road-safety & 100.00\% & 100.00\% \\
        compass & 100.00\% & 100.00\% \\
        \bottomrule
    \end{tabular*}
\end{table}

\section{Binomial vs Hypergeometric Approaches} \label{section:BinomialVsHypergeometric}

Suppose we have an observation $\x$ which we would like to classify using a majority ensemble classifier of the form \eqref{eq:majority_classifier}.
This can be done by tallying the votes $f_1(\x),\, f_2(\x),\,  \dots\,, f_N(\x)$ in sequence, and then computing the majority vote to predict the true class $y$.

In a binary classification context, much of the literature (such as \citet{SchwingZachZhengPollefeys2011} and \citet{SilvaKulldorffKatherineYih2020}) models the total number of observed votes up to $i$ using a binomial distribution $\Binom(i, p)$, under the assumption that the votes are independent of each other, and construct sequential tests on $p$. However, some care must be taken to specify what ``independent" means in this context to determine the validity of this assumption.
The votes $f_1({\x}),\, f_2({\x}),\,  \dots\,, f_N({\x})$ are obviously not \textit{unconditionally} independent, because they all depend on $\x$.
Even if we condition on $\x$, the votes may still be strongly dependent, if they share training data. If they share no training data, and have independent random seeds, then $f_i(\x) \perp f_{i'}(\x) \mid \x$.
Since each vote $f_i(\x)$ is a conditionally independent Bernoulli draw, the number of positive votes $j$, after executing $i$ base models, does indeed distribute as $j \mid \x \sim \Binom(i, p_\x)$, where $p_{\x}$ is the proportion of base models produced from the training distribution that would vote positive on the input $\x$. %While $p_{\x} > 0.5$ is not logically equivalent to $y = 1$ even under this assumption, it is intuitively closely related.

Many ensemble models (including random forests) share some of the training data between base models. When analyzing ensembles of this kind, we must condition both on $\x$ and on the training data $\mathsf{Tr}$ before we can say that the votes are independent of each other. In this case, the parameter $p_\x$ represents the proportion of base models trained on $\mathsf{Tr}$ that would vote positive on the input $\x$, as opposed to the proportion of base models from the training \textit{distribution}. Hypotheses about this $p_\x$ are now one degree further removed from hypotheses about the true label~$y$ due to the possibility of sampling error in $\mathsf{Tr}$, which remains unaccounted for.

For some kinds of ensembles, the independence assumption does not hold even conditioned on both $\x$ and $\mathsf{Tr}$; for example, if the base models are modified to produce certain relationships. An example of such an ensemble is a \textit{pruned} random forest, where individual trees or branches are cut to reduce redundancy.

The approach we take in this paper is to note that, given the total number of positive votes $n$, the number of positive votes after executing $i$ base models distributes hypergeometrically. The hypotheses tested are therefore hypotheses about $n$ rather than about the proportion $p_\x$ of a hypothetical population. The disadvantage of this approach is that hypotheses about $n$ are now even more removed from hypotheses about the true label $y$---not only do we still have unaccounted-for sampling error in $\mathsf{Tr}$, but also in the construction of the base models from $\mathsf{Tr}$. The advantage of our approach is that its guarantees apply even when the weaker independence assumptions are violated, such as in pruned random forests.

\section{Additional Benchmarks} \label{section:BenchmarkResults}

We applied the empirical measurements detailed in Section~\ref{section:EmpiricalResults} to the benchmark datasets proposed by \citet{RandomForestsBetterThanDL} (see their appendix on arXiv for details). Qualitatively, the results are similar to those in the main text, but are presented here for completeness in Figure~\ref{fig:grinsztajn-empirical-performance}.

\begin{figure}[p]
    \includegraphics[width=\linewidth]{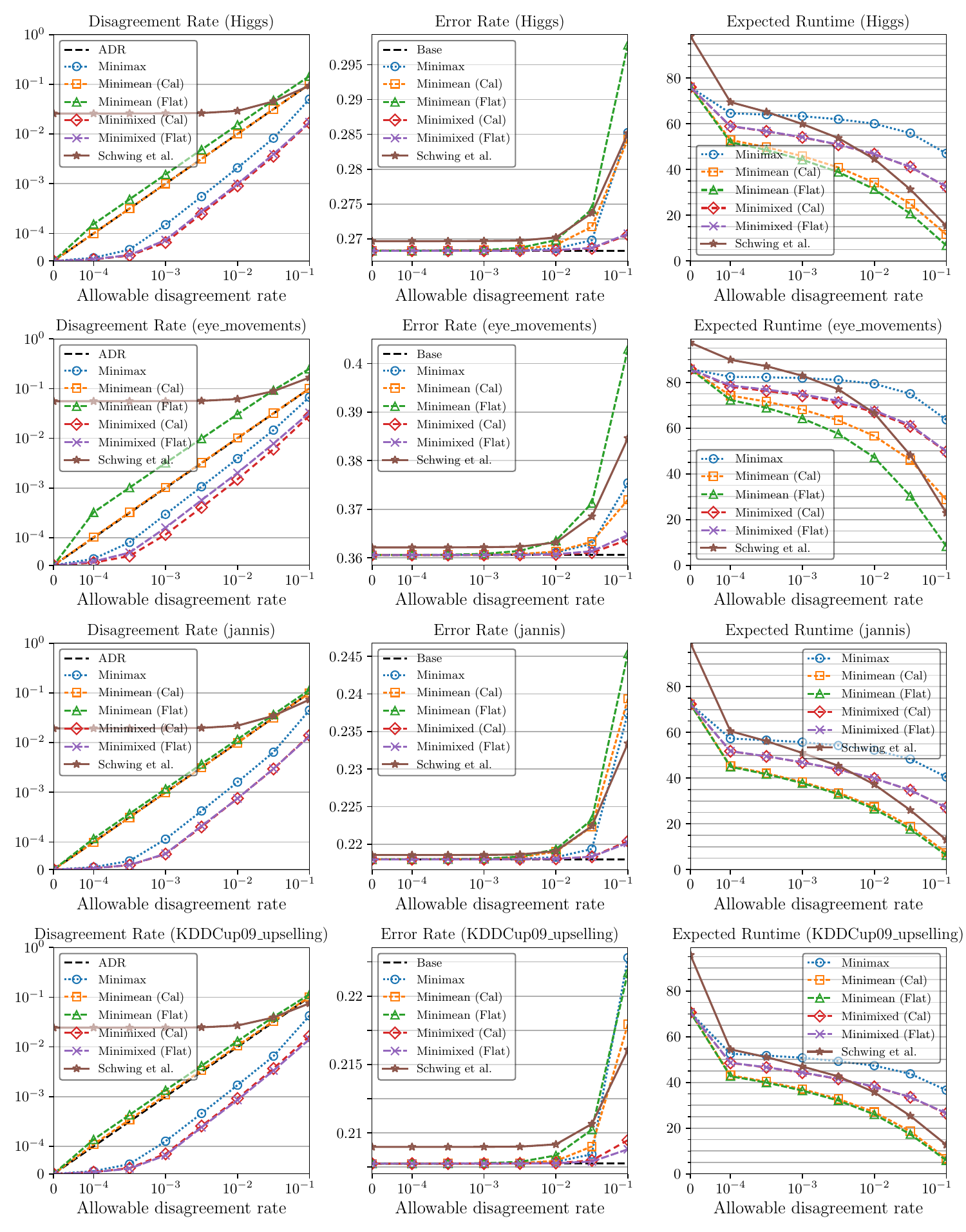}

    \caption{Performance of six early-stopping approaches for random forests of 101 trees on real-world datasets from Grinsztajn et al.'s benchmark. Each row is a dataset, and each column is a different metric. ``ADR" is allowable disagreement rate. ``Expected runtime" means expected number of trees executed before stopping. Page 1 of 5.}
    \label{fig:grinsztajn-empirical-performance}
\end{figure}

\begin{figure}[p]
    \vspace{-3em}

    \ContinuedFloat
    
    \includegraphics[width=\linewidth]{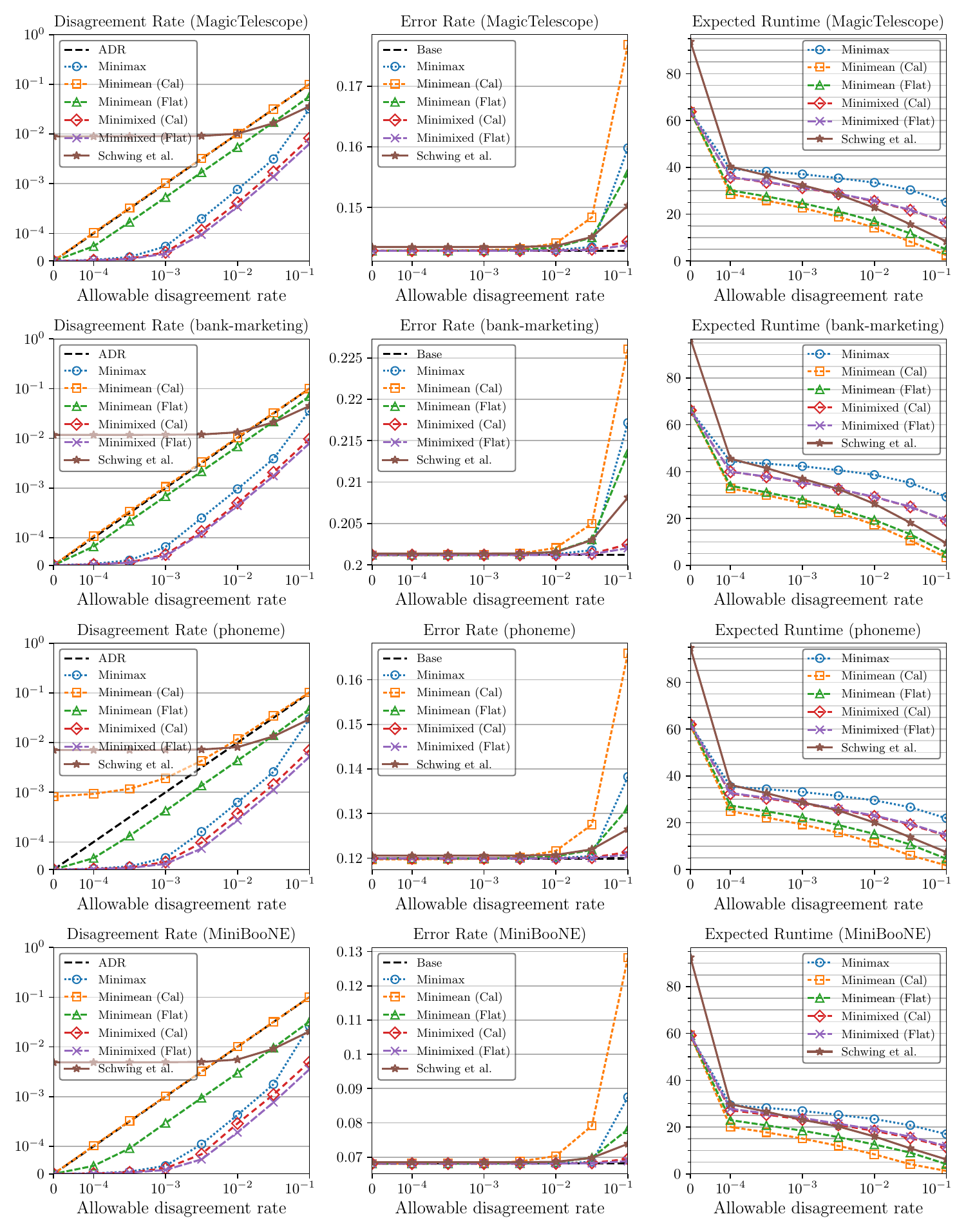}
    
    \caption{Performance of six early-stopping approaches for random forests of 101 trees on real-world datasets from Grinsztajn et al.'s benchmark. Each row is a dataset, and each column is a different metric. ``ADR" is allowable disagreement rate. ``Expected runtime" means expected number of trees executed before stopping. Page 2 of 5.}
\end{figure}

\begin{figure}[p]
    \vspace{-3em}

    \ContinuedFloat
    
    \includegraphics[width=\linewidth]{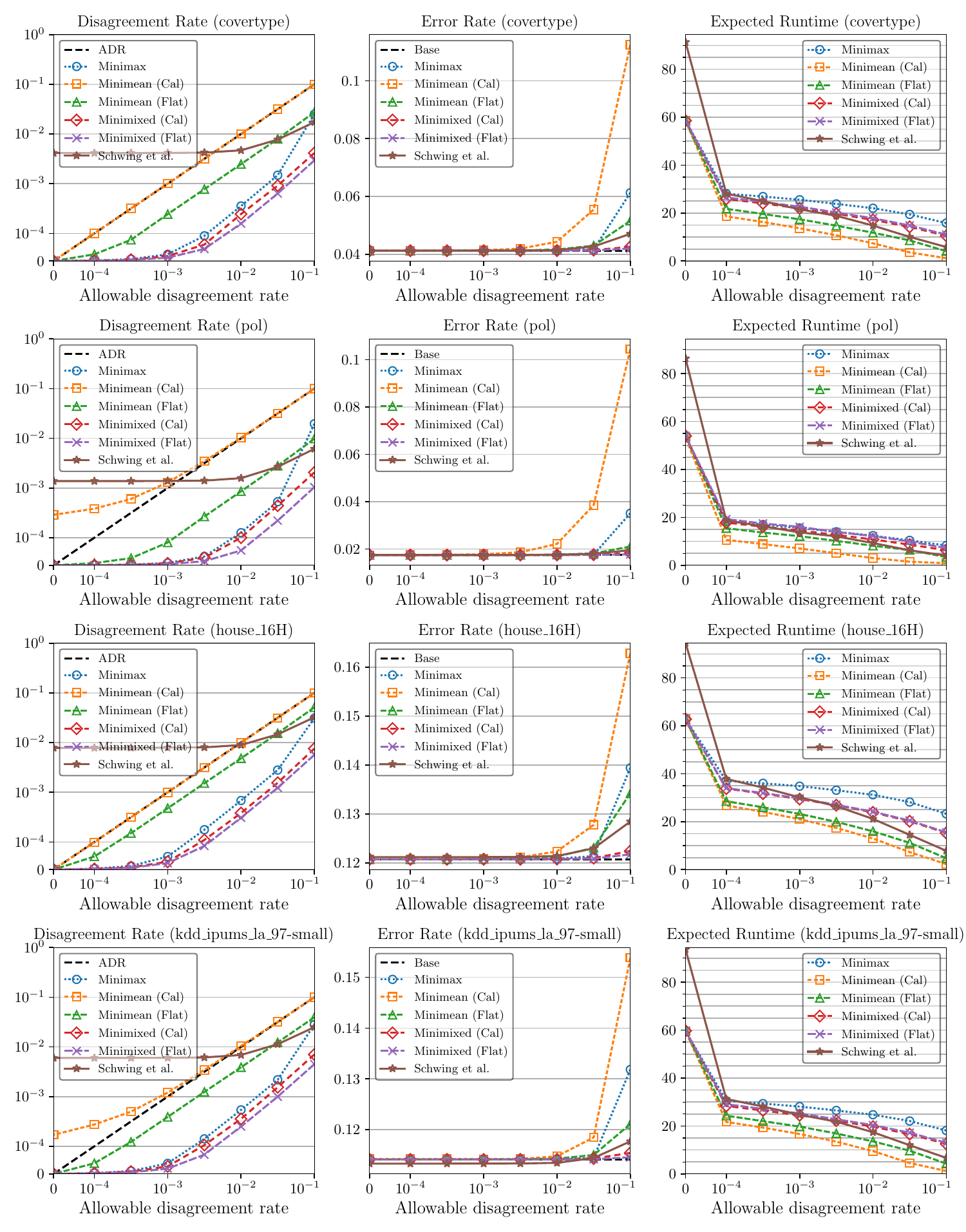}
    
    \caption{Performance of six early-stopping approaches for random forests of 101 trees on real-world datasets from Grinsztajn et al.'s benchmark. Each row is a dataset, and each column is a different metric. ``ADR" is allowable disagreement rate. ``Expected runtime" means expected number of trees executed before stopping. Page 3 of 5.}
\end{figure}

\begin{figure}[p]
    \vspace{-3em}

    \ContinuedFloat
    
    \includegraphics[width=\linewidth]{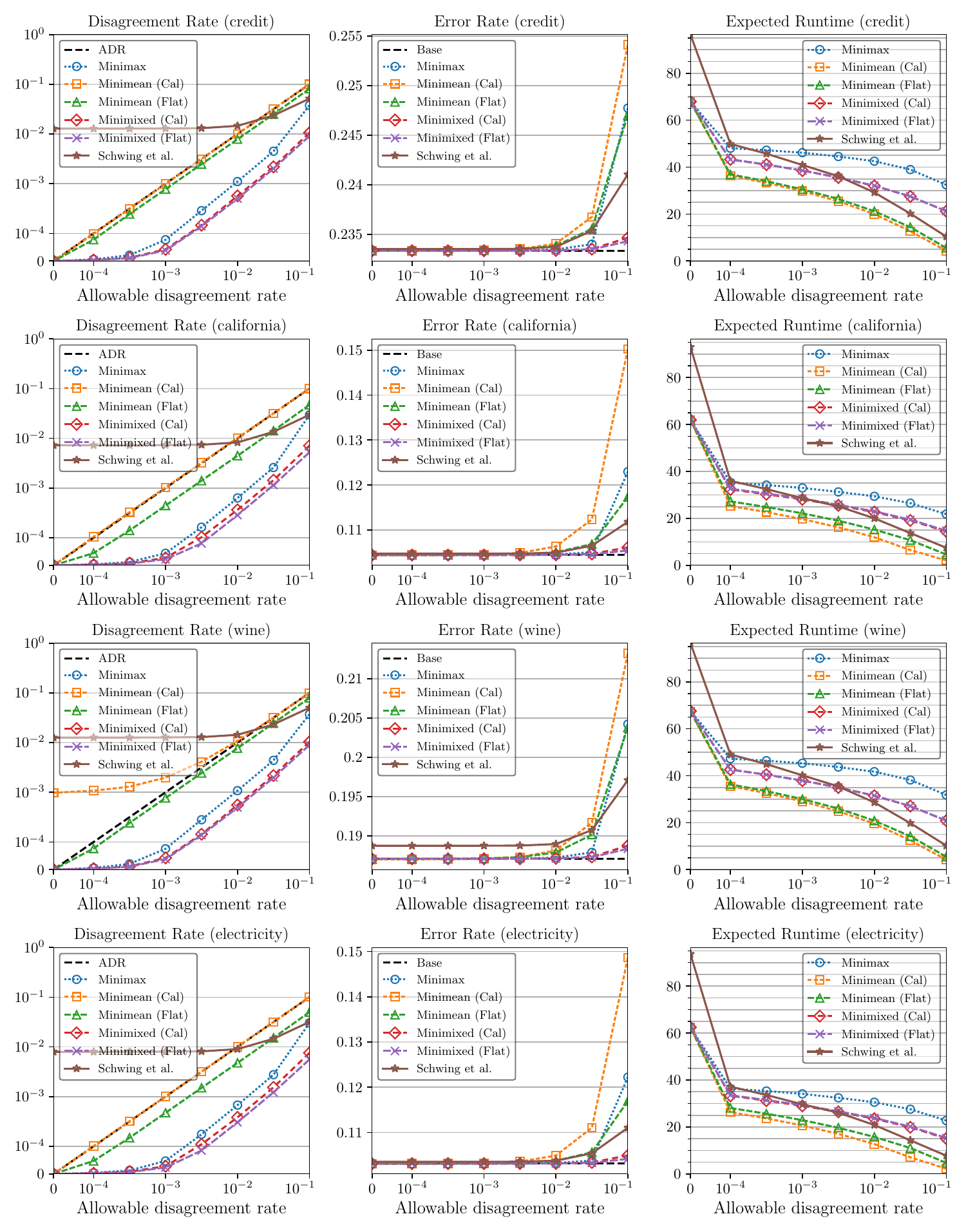}
    
    \caption{Performance of six early-stopping approaches for random forests of 101 trees on real-world datasets from Grinsztajn et al.'s benchmark. Each row is a dataset, and each column is a different metric. ``ADR" is allowable disagreement rate. ``Expected runtime" means expected number of trees executed before stopping. Page 4 of 5.}
\end{figure}

\begin{figure}[p]
    \vspace{-3em}

    \ContinuedFloat
    
    \includegraphics[width=\linewidth]{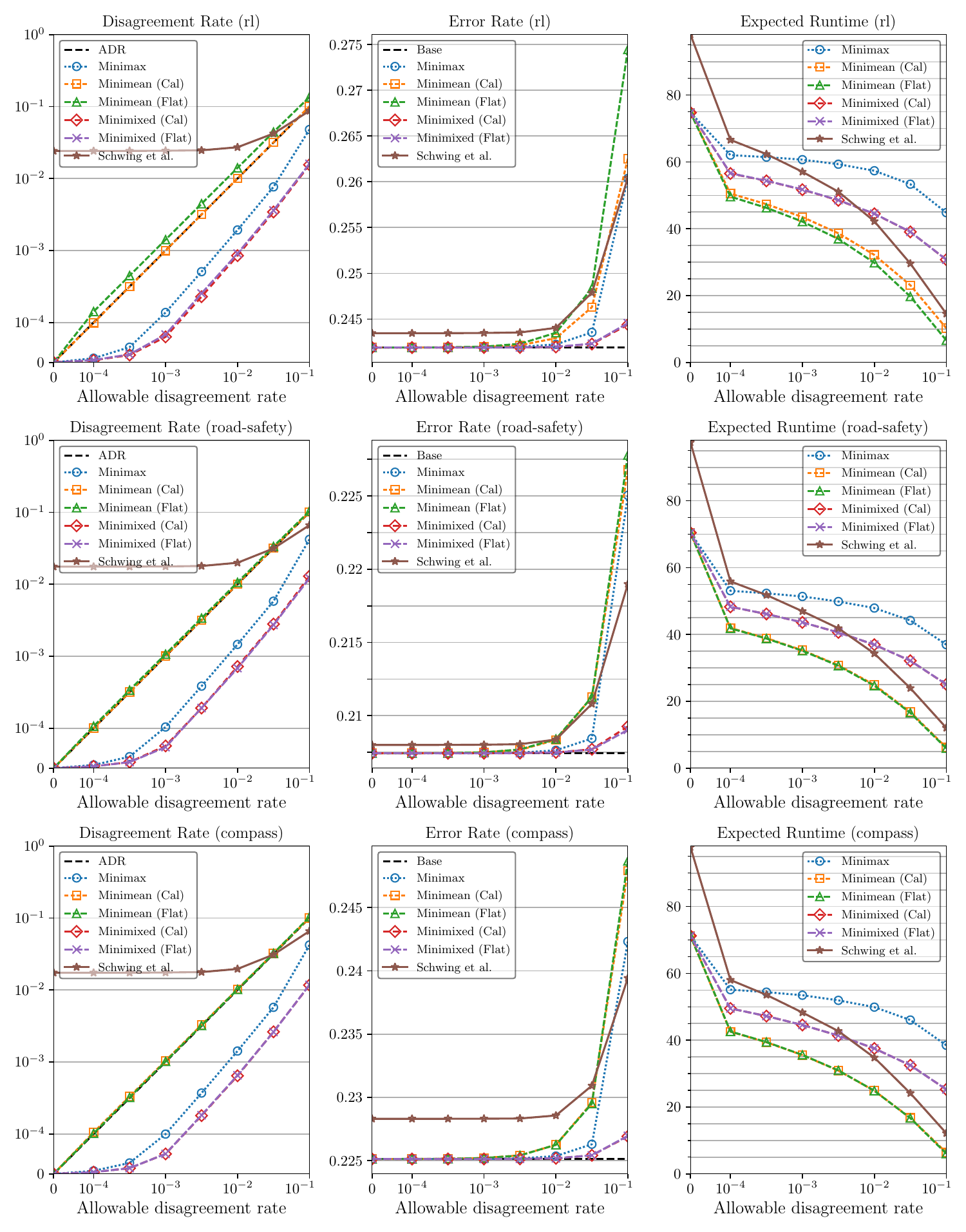}
    
    \caption{Performance of six early-stopping approaches for random forests of 101 trees on real-world datasets from Grinsztajn et al.'s benchmark. Each row is a dataset, and each column is a different metric. ``ADR" is allowable disagreement rate. ``Expected runtime" means expected number of trees executed before stopping. Page 5 of 5.}
\end{figure}

\FloatBarrier

\bibliography{references}
\addcontentsline{toc}{section}{References}